\documentclass[%
  reprint, prl,
superscriptaddress,
 amsmath,amssymb,
 aps,
]{revtex4-2}

\usepackage[
bookmarks=true,
colorlinks,
linkcolor=blue,
urlcolor=blue,
citecolor=olive,
plainpages=false,
pdfpagelabels,
final,
breaklinks=true
]{hyperref}

\usepackage{graphicx}
\usepackage{dcolumn}
\usepackage{bm}
\usepackage{xcolor}
\usepackage{physics}
\usepackage{bbm}
\usepackage{pifont}

\begin{document}

\preprint{APS/123-QED}

\title{Fluctuation-induced symmetry breaking in high harmonic generation for bicircular quantum light}

\author{P.~Stammer}%
\email{philipp.stammer@icfo.eu}
\affiliation{ICFO—Institut de Ciencies Fotoniques, The Barcelona Institute of Science and Technology, 08860 Castelldefels (Barcelona), Spain}%
\affiliation{Atominstitut, Technische Universität Wien, 1020 Vienna, Austria}

\author{C.~Granados}%
\email{cagrabu@eitech.edu.cn}
\affiliation{Eastern Institute of Technology, Ningbo 315200, China}

\author{J.~Rivera-Dean}
\email{physics.jriveradean@proton.me}
\email{javier.dean@ucl.ac.uk}
\affiliation{Department of Physics and Astronomy, University College London, Gower Street, London WC1E 6BT, United Kingdom}

\date{\today}

\begin{abstract}

Symmetries are ubiquitous in physics and play a pivotal role in light-matter interactions, where they determine the selection rules governing allowed atomic transitions and define the associated conserved quantities.~For the up-conversion process of high harmonic generation, the symmetries of the driving field determine the allowed frequencies and the polarization properties of the resulting harmonics.~As a consequence, it is possible to establish classical selection rules when the process is driven by coherent radiation.~In this work, we show that fluctuation-induced symmetry breaking in the driving field leads to the appearance of otherwise forbidden harmonics.~This is achieved by considering bicircular quantum light, and demonstrate that the enhanced quantum fluctuations due to squeezing in the driving field break the classical selection rules.~To this end, we develop a quantum optical description of the dynamical symmetries in the process of high harmonic generation, revealing corrections to the classical selection rules.~Moreover, we show that the new harmonics show squeezing-like signatures in their photon statistics, allowing them to be clearly distinguished from classical thermal fluctuations.
\end{abstract}

\maketitle


Symmetries are fundamental in physics due to their connection to conservation laws, as established by Noether~\cite{noether1983invariante}.~In the context of light–matter interaction, symmetries are ubiquitous: they underlie the minimal-coupling Hamiltonian implied by gauge invariance~\cite{cohen2024photons}, and govern optical transitions within Fermi’s golden rule~\cite{dirac1927quantum, fermi1950nuclear}.~More generally, selection rules encode symmetry constraints on transitions between quantum states and provide direct physical insight in terms of conserved quantities such as energy, parity, and angular momentum.~For instance, in electric-dipole transitions in isotropic media, parity must change between the initial and final states, while angular momentum follows the familiar dipole selection rules~\cite{cohen2024atom}.

In strong-field-driven processes, particularly high-harmonic generation (HHG), conservation of energy and parity implies that, for a monochromatic, linearly polarized driving field in isotropic media, only odd harmonics $q = 2n+1$ are emitted.~From the perspective of the symmetries of the driving laser, the appearance of odd harmonics and their polarization originates from invariance under the joint operation of a temporal translation $\hat{\tau}_2 = \pi/\omega$, and a spatial rotation $\hat{r}_2 = \pi$. More generally, the theory of dynamical symmetries provides a framework to derive selection rules governing the harmonic spectrum~\cite{alon1998selection}, with applications ranging from solids~\cite{he2022dynamical} to topological materials~\cite{ma2022role}. Within this framework, the symmetries of the classical driving field are mapped onto observable properties of the emitted radiation, such as the allowed harmonic orders, their polarization state, and helicity~\cite{neufeld2025light,Nariyuki_SLSolids,Tang_SL, Nimrod_SL, Naotaka_Graphene}.~Dynamical symmetries thus provide a unified description for classifying harmonic spectra and identifying emission channels that are allowed or forbidden by symmetry~\cite{lerner_multiscale_2023}. They also explain selection rules for more complex driving fields, such as bichromatic circularly polarized fields~\cite{baykusheva2016bicircular}. Counter-rotating bicircular fields (BCF) are of particular interest, as their symmetries give rise to characteristic selection rules in isotropic media, where harmonics at orders $q = 3n \pm 1$ are allowed and exhibit alternating helicity, while the $q = 3n$ harmonics are symmetry forbidden~\cite{pisanty_spin_2014,fleischer_spin_2014}. 

Therefore, extending classical bicircular fields by including their quantum nature is particularly attractive to understand the role of non-classical fluctuations on the symmetries of the light-matter interaction, and how those fluctuations allow control of the classical selection rules (CSR).  
Nonetheless, all previous studies of selection rules in HHG and their dynamical-symmetry origins have been restricted to classical driving fields. In this case, the field is treated deterministically, with well-defined amplitude, phase, and symmetry properties. Consequently, the aforementioned selection rules strictly hold.~Quantum optics, however, enables the inclusion of nonclassical driving fields, such as squeezed states of light~\cite{loudon1987squeezed}, into the HHG light–matter interaction problem, thereby opening the door to explore the impact of quantum fluctuations of the driving field on symmetry-related effects.

\begin{figure}
    \centering
    \includegraphics[width=1\columnwidth]{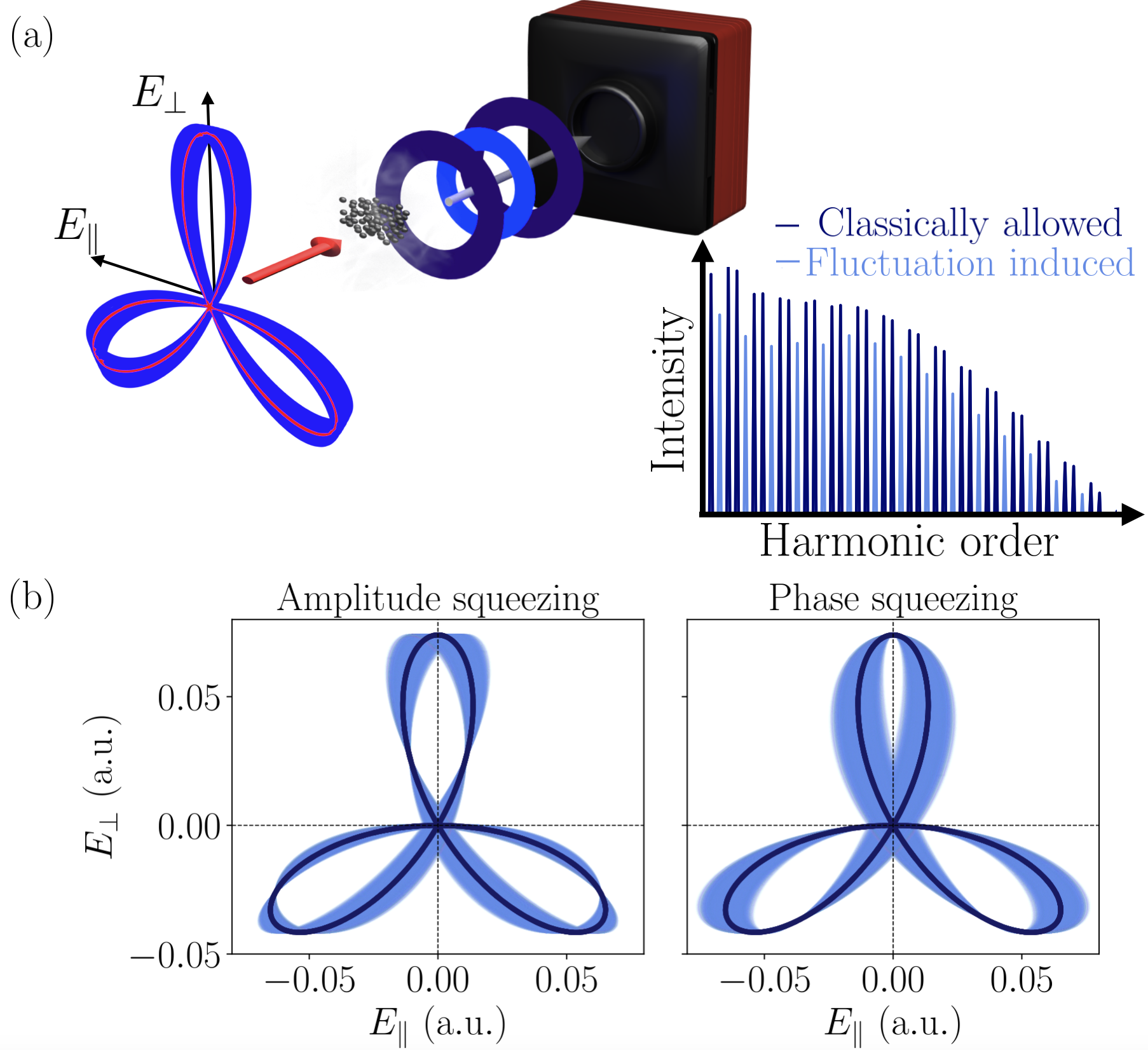}
    \caption{(a) Schematic figure of BQL leading to symmetry breaking in the field, and hence the observation of classically forbidden harmonics induced by the quantum fluctuations of the field.
    (b) Lissajous figures of the BQL for amplitude and phase squeezed $2 \omega$ field. The dark blue lines show the mean field value, following the dynamical symmetry, while the shaded region indicates the field fluctuations breaking the symmetry.}
    \label{Fig:lissajous}
\end{figure}

In this work, we demonstrate that the well-known classical bicircular selection rules are modified in the presence of nonclassical driving fields through symmetry breaking of the system.~We show that this symmetry breaking is induced by quantum fluctuations originating from the squeezing of the fundamental BCF. Extending classical bicircular selection rules to include quantum light connects naturally with recent developments in strong-field quantum optics~\cite{lewenstein2021generation, stammer2025colloquium, stammer2025theory}, and in particular with the growing interest in HHG driven by quantum light~\cite{gorlach2023high, rasputnyi2024high,lemieux_photon_2025,tzur_measuring_2025, gothelf2025high, rivera2025structured, petrovic_generation_2026, wang2025high}.~Our results place the concept of dynamical symmetries within a genuine quantum optical framework and establish how symmetry breaking emerges when HHG is driven by bicircular quantum light (see Fig.~\ref{Fig:lissajous}~(a)).


\emph{Framework.--}  
Since we are interested in driving the process of HHG by bicircular quantum light (BCL), we start by describing the classical counterpart. A bicircular field is a bichromatic field consisting of a fundamental mode $\omega$, and its second harmonic $2 \omega$ of equal amplitude but opposite helicity. 
The overlap of the left and right circular polarized fields results in a trefoil-shaped total electric field (see Fig.~\ref{Fig:lissajous}), having a rotational symmetry by $2 \pi/3$ in polarization. Therefore, a classical bicircular field is described by the electric field components in the polarization plane as $E_\parallel(t) = -E_0[\sin(\omega t)+\sin(2\omega t)] $ and $E_\perp (t) = E_0[\cos(\omega t)-\cos(2\omega t)]$, as illustrated in Fig.~\ref{Fig:lissajous}~(b) by the dark blue classical field average. 

Before addressing the consequences of BQL on light–matter interactions, we briefly recall the selection rules for HHG driven by classical bicircular fields.~In contrast to single-color, linearly polarized drivers, which generate only odd, linearly polarized harmonics $q = 2n+1$, $n \in \mathbb{N}$, bicircular fields yield harmonics governed by symmetry-imposed selection rules
\begin{align}
\label{eq:CSR}
    q = 3n \pm 1, \quad n \in\mathbb{N},
\end{align}
while the $q = 3n$ harmonics are forbidden by the underlying dynamical symmetry.~The allowed harmonics possess well-defined polarization states: the $q = 3n+1$ and $q = 3n-1$ orders carry the same helicity as the $\omega$ and $2\omega$ driving components, respectively. This structure can be understood from angular momentum conservation, expressed in terms of the number of absorbed photons from each mode, $n_1$ (for $\omega$) and $n_2$ (for $2\omega$), which satisfy $n_2 - n_1 = \pm 1$~\cite{pisanty_spin_2014,fleischer_spin_2014}. This constraint directly reflects the threefold dynamical symmetry of the bicircular driving field~\cite{pisanty_spin_2014}.

The quantum optical counterpart of the classical bicircular field is represented by a product of coherent states in the $\omega$ and $2\omega$ mode $\ket*{\alpha_{\omega, L}} \otimes \ket*{\alpha_{2\omega, R}}$, having left ($L$) and right ($R$) circular polarization, respectively.
Here, we consider the scenario of BQL in which one of the field components exhibits squeezing signatures. Specifically, we consider the case in which we drive the process of HHG by the initial state with squeezing along the parallel $2 \omega$-component (see Supplementary Material for alternative configurations)
\begin{equation}
\label{eq:state_initial}
	\ket{\Phi(t_0)}
		= \hat{D}_{\omega,L}(\alpha_\omega\sqrt{2})
				\otimes [\hat{D}_{2\omega,R}(\alpha_{2\omega}\sqrt{2})\hat{S}_{2\omega,\parallel}(\xi)]\ket{\bar{0}},
\end{equation}
where $\hat{D}_{\omega,\mu}(\alpha)$ and $\hat{S}_{\omega,\mu}(\xi)$ are the displacement and squeezing operators acting on the optical mode $(\omega,\mu)$, respectively.~Furthermore, we defined the overall vacuum state $\ket{\bar{0}}= \bigotimes_{q} \ket{0_q}$.~To ensure that in the absence of squeezing, $\xi=0$, the driving field exhibits the correct dynamical symmetry of a classical bicircular field, we have $\abs*{\alpha_{\omega}} = \abs*{\alpha_{2\omega}}$.~Since we can express the polarization state in any basis, we shall decompose Eq.~\eqref{eq:state_initial} in the linear basis given by the horizontal and vertical polarization component, such that $\hat{D}_{R}(\alpha\sqrt{2}) \equiv \hat{D}_{\parallel}(\alpha) \otimes \hat{D}_{\perp}(i \alpha)$, and $\hat{D}_{L}(\alpha\sqrt{2}) \equiv \hat{D}_{\parallel}(\alpha) \otimes \hat{D}_{\perp}(-i \alpha)$.~Hence, the initial $2 \omega$ field can be written as
\begin{equation}
    \begin{aligned}
        \ket{\Phi_{2 \omega}(t_0)} &=  \hat{D}_{2\omega,\parallel}(\alpha_{2\omega}) \hat{S}_{2\omega,\parallel}(r) \ket{0_\parallel}
            \\ &\quad \otimes \hat{D}_{2\omega,\perp}(i \alpha_{2\omega}) \ket{0_\perp}.    
    \end{aligned}
\end{equation}

An example of the BQL is shown in Fig.~\ref{Fig:lissajous}~(b) by means of the Lissajous figure of the field for amplitude and phase squeezing in the perpendicular $2 \omega$ component. In the figures, the average field (dark blue) follows the dynamical symmetry by rotation of $2 \pi/3$, while the fluctuations (shaded region) clearly break the symmetry. In the following, we will demonstrate how this fluctuation-induced symmetry breaking leads to the appearance of classically forbidden harmonics in the HHG process. 

To simulate the HHG process driven by BQL, we consider the electron initially in the ground state and solve the Schrödinger equation within the strong-field approximation under the full quantum optical Hamiltonian in the dipole coupling~\cite{stammer2023quantum}, with the electric field operator defined as
\begin{equation}
\label{eq:field_operator}
    \hat{\boldsymbol{E}}(t) = - i \!\!\sum_{\mu = R,L} \!\sum_{q}\kappa_q \boldsymbol{\epsilon}_\mu  \left[ \hat{a}_{q,\mu} \,e^{-i\omega_q t} +  \hat{a}^\dagger_{q,\mu} \, e^{i\omega_q t} \right],    
\end{equation}
where $\boldsymbol{\epsilon}_\mu$ denotes the polarization vector and $\kappa_q$ the coupling constant. 
Given the techniques developed in previous work~\cite{rivera-dean_attosecond_2025,petrovic_generation_2026}, we can solve the dynamics in the strong field regime~\cite{lewenstein2021generation, stammer2023quantum, rivera2022strong, stammer2024entanglement}, such that the final state of the $q$th harmonic mode is given by (see End Matter)
\begin{equation}
\label{eq:harmonic_state}
	\hat{\rho}_{q,\mu}(t)=\int \dd^2 \alpha\ Q(\alpha) \dyad{\chi_{q,\mu}(\alpha)}{\chi_{q,\mu}(\alpha)},
\end{equation}
where $Q(\alpha) = \pi^{-1}\rvert\langle\alpha\vert\Phi_{2\omega,\parallel}(t_0)\rangle\rvert^2$ represents the Husimi function of the squeezed polarization component in the driving field.~The amplitude $\chi_{q,\mu}(t) = \text{FT}_q[\langle \hat{d}_{\mu}(t;\alpha)\rangle]$ is the Fourier transform of the dipole moment obtained from the solution of the semi-classical Schrödinger equation driven by a classical field of a coherent state $\ket{\alpha}$~\cite{RBSFA}.
With the state of the harmonics in Eq.~\eqref{eq:harmonic_state} we can now obtain all properties and observables of the harmonic radiation. 

\begin{figure}
    \centering
    \includegraphics[width=1\columnwidth]{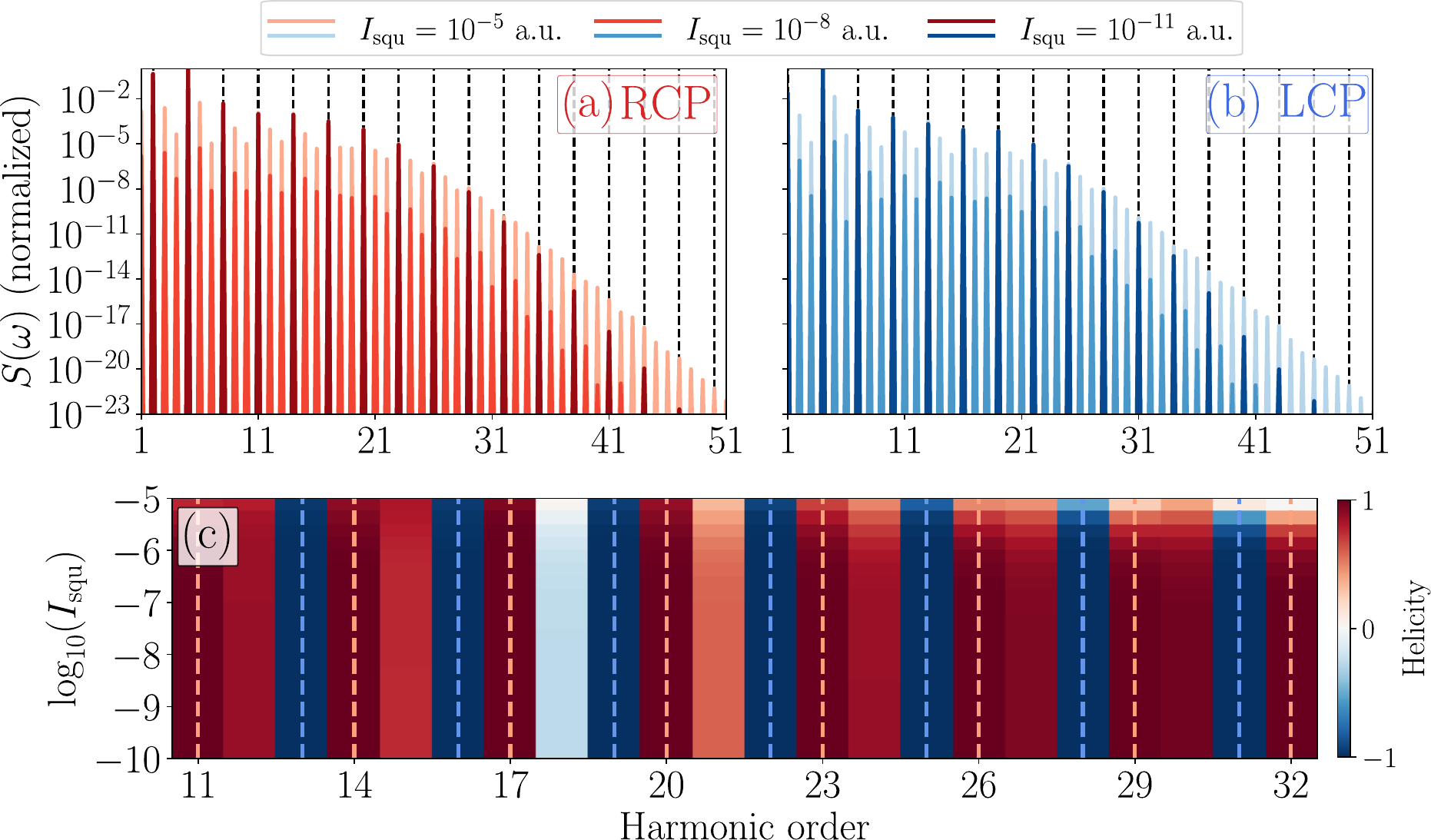}
    \caption{
    (a), (b) HHG spectra resolved along the right- and left-circular polarization components, respectively.~The black dashed lines indicate the positions of the classically allowed harmonics for each polarization component:~(a) $3n-1$ and (b) $3n + 1$.~(c) Helicity as a function of the squeezing intensity, with the dashed lines highlighting the classically allowed harmonics.~The atomic medium is hydrogen ($I_p = 0.5$ a.u.), the driving frequency is $\omega_L = 0.057$ a.u., and the bicircular coherent field amplitudes are set to $E_0 = 0.037$ a.u.
    }
    \label{Fig:HHG:spec}
\end{figure}

\emph{Fluctuation-induced harmonics.--} 
First, we want to understand how the presence of the non-classical signatures in the BCL affects the harmonic spectrum. For the case of squeezing in the parallel component of the $2 \omega$ field, the HHG spectrum from the BCL for different squeezing intensity is shown in Fig.~\ref{Fig:HHG:spec}, split into the $R$-circular (RCP) and $L$-circular (LCP) polarized harmonics in panel (a) and (b), respectively.
The dark colored harmonic peaks in both spectra indicate a vanishing squeezing intensity, effectively reproducing the classical harmonic spectra from the conventional bicircular driving field (also indicated by the dashed vertical lines at the classically allowed harmonics $q = 3n \pm 1$).

In striking difference, as the squeezing strength increases, harmonics that are forbidden in the classical case ($q = 3n$) emerge in the spectrum. Notably, these newly generated harmonics contain both RCP and LCP components. Moreover, the classically allowed harmonics in Eq.~\eqref{eq:CSR}, develop additional contributions from the conjugate polarization state in the presence of squeezing.
This has the interesting consequence that the classically $R$ polarized harmonics have an additional contribution from the conjugate $L$ polarization when driven by quantum light, and vice versa.~These results demonstrate that quantum fluctuations not only break the underlying dynamical symmetry, but also enable control over the helicity and intensity of the $q = 3n$ harmonic orders.~This modification is further illustrated in Fig.~\ref{Fig:HHG:spec}(c), where we show the helicity of the harmonics as a function of the squeezing strength. The harmonics obeying Eq.~\eqref{eq:CSR} largely retain their original polarization, as the contribution from the opposite helicity remains small. In contrast, the classically forbidden harmonics ($q = 3n$) exhibit a dominant RCP component, i.e. they inherit the polarization of the driving field mode in which squeezing is applied (see SM for details). This behavior indicates that these classical forbidden harmonics originate from the squeezed component of the field, and directly reflect its origin from the quantum fluctuations.
Note that only for the harmonics beyond the cut-off, and at very large squeezing intensities, the effective polarization of the harmonics becomes linear.

\emph{Breaking classical selection rules.--} What is the origin of this violation of the classical selection rules (CSR), and why do classically forbidden harmonics appear? We show that this behavior arises from the breaking of the underlying dynamical symmetry of the driving field.~For a classical bicircular field, the relevant dynamical symmetries are discrete rotations and time translations, described by the operators $\hat r_n$ and $\hat \tau_n$, corresponding to a rotation of the field polarization by $2 \pi/n$ and a temporal shift of $T/n$, respectively.~Here, $T = 2 \pi / \omega$ is the period of the fundamental field.~Looking at the 3-fold symmetry of the classical field in Fig.~\ref{Fig:lissajous}, we identify that the bicircular field is invariant under the joint symmetry operation of $\hat r_3$ and $\hat \tau_3$, leading to the classical selection rule of Eq.~\eqref{eq:CSR}, whereas the harmonics $q = 3n$ are symmetry forbidden.~While the symmetry operations $\hat r_n$ and $\hat \tau_n$ act on the classical fields, i.e. $\hat r_n \hat \tau_n \vb{E}_\mu(\vb{r},t) \boldsymbol{\varepsilon}_{\perp}  = \vb{E}(R(\theta_n)\vb{r}, t+ T/n)  = \vb{E}(\vb{r}, t+ T/n) [\cos(2\pi/n)\boldsymbol{\varepsilon}_{\perp}+\sin(2\pi/n)\boldsymbol{\varepsilon}_{\parallel}]$, where $R(\theta_n)$ is the rotation matrix by the angle $\theta_n = 2 \pi/n$. 

In contrast, in this work we are concerned with quantum optical fields. Consequently, we define the quantum optical symmetries of the field and its fluctuations by using the unitaries leading to a rotation in polarization 
\begin{equation}\label{Eq:rotation}
    \hat{R}(\theta) = \exp 
    \left[
    -i\theta
    \left(
    \hat{a}^\dagger_R \hat{a}_R - \hat{a}^\dagger_L \hat{a}_L
    \right)
    \right],
\end{equation}
and the unitary for temporal translation in the $R/L$ basis 
\begin{equation}\label{Eq:time:translation}
    \hat{T}(\theta) = \exp
    \left[
    -i\theta
    \left(
    \hat{a}^\dagger_R \hat{a}_R + \hat{a}^\dagger_L \hat{a}_L
    \right)
    \right],
\end{equation}
corresponding to the propagation under the free-field Hamiltonian for a time $\theta  = \omega t$. 
Now, we proceed to understand the role of the dynamical symmetry transformation in the quantum optical scenario by considering the electric field operator from Eq.~\eqref{eq:field_operator}, instead of the classical field as previously done~\cite{neufeld2025light}. 
Therefore, we define the electric field operator after acting with the unitary symmetry operators 
\begin{align}
    \hat{E}(t; \theta, \tau) = \hat{T}^\dagger(\omega \tau)\hat{R}^\dagger(\theta) \hat{E}(t) \hat{R}(\theta) \hat{T}(\omega \tau).
\end{align}

With the symmetry transformed electric field operator, we are able to understand how the quantum bicircular light breaks the CSR. First, we note that the average field $\expval*{\hat{E}(t; \theta, \tau)} = E_{\text{cl}}(t;\theta, \tau)$, for the initial state of Eq.~\eqref{eq:state_initial}, obeys the same symmetry as the classical bicircular field
\begin{align}
    \expval{\hat{E} \left( t; \frac{2 \pi}{3}, \frac{2 \pi}{3 \omega} \right) } = \expval{\hat{E}(t)} = E_{\text{cl}}(t),
\end{align}
for a rotation and time-translation by $2 \pi/3$ and $2 \pi/(3 \omega)$, respectively.~Thus, on average, the quantum bicircular field obeys the same dynamical symmetry as the classical bicircular field.~However, the striking difference lies in the quantum fluctuations of the field due to the squeezing contribution, which do not follow the classical symmetry.~In particular, the variance of the electric field $(\Delta E)^2(t;\theta, \tau) = \expval*{\hat{E}^2(t;\theta, \tau)} - \expval*{\hat{E}(t;\theta, \tau)}^2$ is given by (see End Matter)
\begin{equation}\label{Eq:variance}
\begin{aligned}
    (\Delta E)^2 & (t;\theta, \tau)  = \cosh(r) \sinh(r) \\ 
     & \quad  \times \left[ 4 \coth(r)  - \cos(4 \omega (t+\tau)) \right] .
\end{aligned}
\end{equation}

While the average field of the quantum state in Eq.~\eqref{eq:state_initial} preserves the symmetry, the field fluctuations generally break the dynamical symmetry. The stronger the squeezing amplitude $r = \abs*{\xi}$, the larger the field fluctuations, and hence the breaking of the dynamical symmetry is more pronounced. This has the interesting consequence that the harmonic response to the BQL is no longer constrained by the symmetry of the classical mean field alone, but the fluctuation-induced symmetry breaking opens additional channels for the emission of harmonic photons. These new harmonics, as seen in the spectra of Fig.~\ref{Fig:HHG:spec}, are forbidden within the purely classical description.
Given the quantum optical framework introduced here, we can now write the origin of the fluctuation-induced harmonics in terms of modified selection rules. For this, we express the linearly polarized squeezing contribution in the $R/L$-circularly polarized basis. Therefore we define the number of photons absorbed from the $\omega$ mode, $n_{1,-}$, and the coherent and squeezed $2\omega$ photon numbers $n_{2,+}$ and $n_{2,-}$, respectively. The index $+/-$ indicates the $R/L$ polarization of the driving field modes. 
Hence, the new selection rule is given by (see End Matter)
\begin{equation}\label{Eq:new:harm:rule}
	q = n_{1,-} + 2 \, ( n_{2,+} + s \, n_{2,-}) ,	
\end{equation}
where $s = 1$ if there is squeezing in the driving field, and $s=0$ in the classical case without squeezing. Furthermore, the number of absorbed photons is restricted by the angular momentum
\begin{equation}\label{Eq:new:spin:rule}
    \sigma_{q} = n_{2,+} - ( s \, n_{2,-} + n_{1,-} ),
\end{equation}
which naturally opens two new absorption channels determining by the polarization of the harmonics, $\sigma_q=\pm1$ (see the SM for detailed calculations).~Note that the classical selection rules are recovered when $s=0$, and hold true for isotropic HHG media such as atoms, and with the interaction in the dipole approximation.~Other media, such as molecules~\cite{ceccherini2001dynamical} or solids provide different selection rules~\cite{Nariyuki_SLSolids,
Naotaka_Graphene}, and higher order couplings can lead to different symmetry properties~\cite{lysne2020high, kong2019spin}.~Having established the quantum optical formulation of the dynamical symmetry for HHG, and shown its violation for quantum light, we now continue to investigate the photon statistics of these classical forbidden harmonics.

\begin{figure}
    \centering
    \includegraphics[width=1\columnwidth]{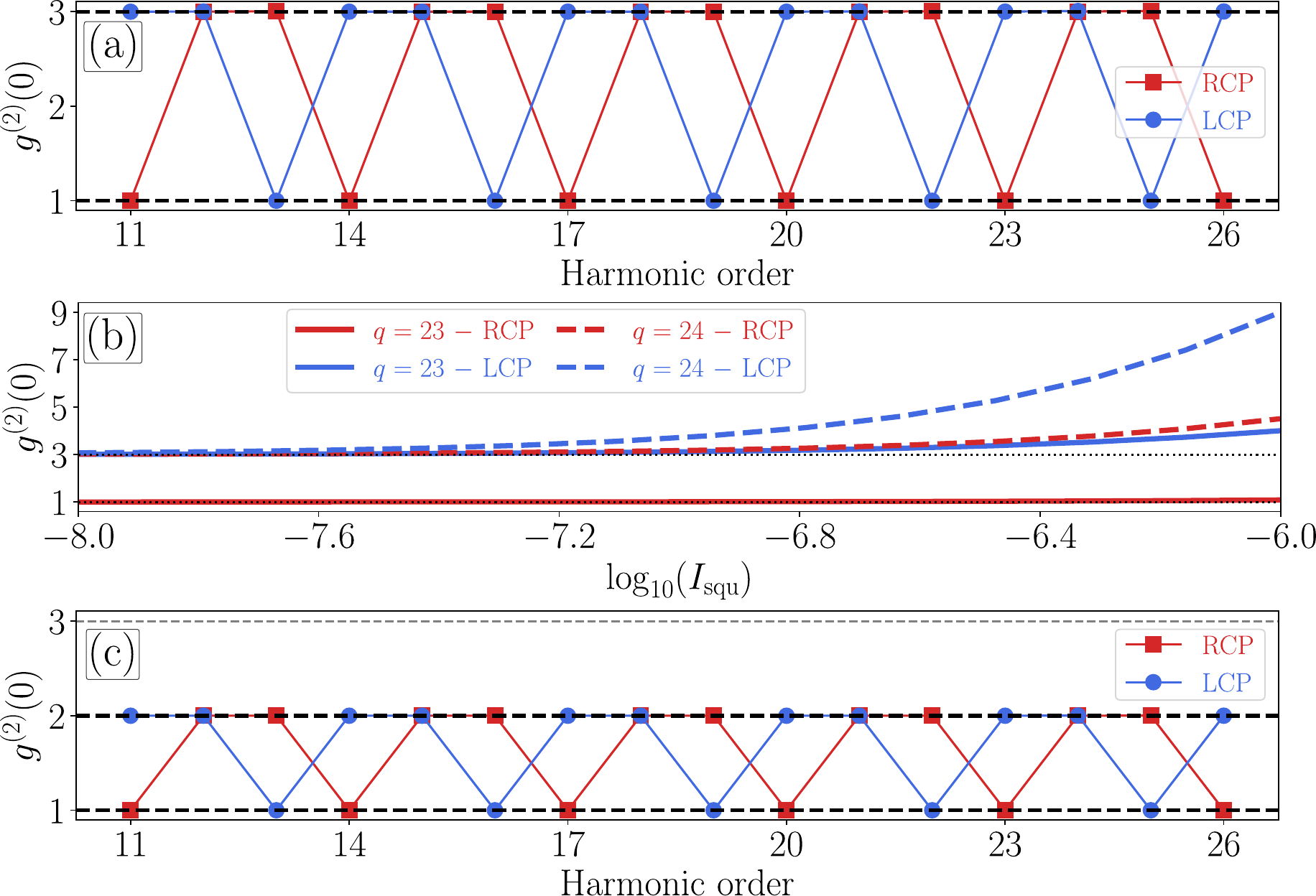}
    \caption{Equal time intensity correlation function $g^{(2)}(0)$ for a squeezed driver [(a),(b)] and a thermal state [(c)].~Panel (a) shows the $g^{(2)}(0)$ for different harmonic orders, resolved into $R$- and $L$-polarization components (red and blue, respectively), for a squeezing intensity $I_{\text{squ}} = 10^{-8}$ a.u.~Panel (b) displays $g^{(2)}(0)$ for two selected harmonic orders as a function of the squeezing intensity. Panel (c) is analogous to (a), but for a thermal state with $I_{\text{th}} = 10^{-9}$ a.u. The atomic medium is hydrogen ($I_p = 0.5$ a.u.), the driving frequency is $\omega_L = 0.057$ a.u., and the bicircular coherent field amplitudes are set to $E_0 = 0.037$ a.u.}
    \label{Fig:g2:squz}
\end{figure}

\emph{Fluctuation-induced photon statistics.--}~We have clearly demonstrated that the appearance of forbidden harmonics originate from the violation of the classical dynamical symmetry in the field.~However, it is important to establish the statistics of the emitted photons, and whether they inherit the properties of the quantum driving field.~To investigate this, in Fig.~\ref{Fig:g2:squz} we show the equal time intensity correlation function for the generated harmonics with their respective polarization (see End Matter and Refs.~\cite{petrovic_generation_2026, rivera-dean_attosecond_2025, stammer2025weak})
\begin{equation}\label{Eq:SM:g2}
	g_{q,\mu}^{(2)}(0)
		= \dfrac{\langle \hat{a}_{q,\mu}^{\dagger 2}\hat{a}_{q,\mu}^2\rangle}{\langle \hat{a}_{q,\mu}\hat{a}_{q,\mu}\rangle^2}
		= \dfrac{\int \dd^2 \alpha\ Q(\alpha) \abs{\chi_{q,\mu}(\alpha)}^4}{\big[\int \dd^2 \alpha \ Q(\alpha)  \abs{\chi_{q,\mu}(\alpha)}^2\big]^2}.
\end{equation}

In panel Fig.~\ref{Fig:g2:squz}~(a) we can see that the $R$ ($L$) polarized harmonics have a classical Poisson distribution of $g^{(2)}(0) =1$ at their classical allowed harmonic orders $q = 3n - 1$ ($q = 3n +1$).~In contrast, all harmonics which appear due to breaking the classical selection rule have a photon number distribution indicating super-Poissonian statistics, just like that of a squeezed vacuum field with $g^{(2)}(0) = 3$.~This implies that the classically forbidden harmonics at $q=3n$ have squeezed-like photon statistics for both $R$ and $L$ components.~For the classically allowed harmonics, the same super-Poissonian $g^{(2)}(0)=3$ value is obtained when analyzed along the polarization opposite to that in which they are generated.~In contrast, when measured in their native (classically allowed) polarization, these harmonics recover Poissonian statistics.
Hence, we have seen that these fluctuation-induced harmonics show super-Poissonian statistics, originating from the squeezing in the driving field.~Now, increasing the squeezing intensity we show in Fig.~\ref{Fig:g2:squz}~(b) the $g^{(2)}(0)$ function for exemplary harmonics as the squeezing amplitude increases.~In particular, we show the classical allowed $q_R = 23$ harmonic with $R$-polarization (solid red), which shows Poissonian distribution $g^{(2)}(0)=1$ independent of the squeezing intensity. In contrast, the classical forbidden harmonics, namely the $q_L=23$ with $L$-polarization (solid blue), as well as the forbidden $q=24$ ($R$ and $L$, dashed), are increasing for larger squeezing.
Interestingly, the symmetry forbidden harmonics $q=24$ scale stronger with the increased squeezing, further manifesting their origin purely from the quantum fluctuations of the field. 

However, one may wonder if the observations made thus far are due to genuine quantum fluctuations, or if classical fluctuations would give similar results.~Therefore, we also consider the case of thermal fluctuations instead of squeezing (see End Matter for details). While the classical fluctuations of the thermal light also break the symmetry of the field, the influence on the $g^{(2)}(0)$ function remains trivial, as shown in Fig.~\ref{Fig:g2:squz}~(c).~Although the alternating polarization property of the harmonics persists, the photon statistics of the symmetry forbidden harmonics merely displays that of thermal fluctuations $g^{(2)}(0)=2$. 
Hence, the measurement of the $g^{(2)}(0)$ function allows to unambiguously deduce the presence of genuine quantum fluctuations of the symmetry forbidden harmonics.

\emph{Conclusion.--}
In this work, we have established the quantum optical framework for dynamical symmetries in HHG driven by quantum light.~We have demonstrated that it is insufficient to consider only the classical mean of the quantum optical field, as the intrinsic quantum fluctuations break the underlying classical symmetry.
This manifests in the appearance of symmetry-forbidden harmonics in the spectrum, thereby violating classical selection rules, and leaves squeezing-like imprints in the photon statistics of these symmetry-forbidden harmonics.

By introducing the symmetry operations that act directly on the field operator, we generalized the concept of dynamical symmetries beyond the classical description, and extend the classical selection rules to account for the quantum fluctuations governing quantum optical HHG.~Given the generality of dynamical symmetries and selection rules, the framework established here opens new avenues for investigating the role of symmetries in the quantum optical context.~It remains an open question how the quantum state of the symmetry-broken harmonics is affected by these fluctuations, which is closely related to information theoretic approaches where conservation laws impose constraints directly on the quantum state~\cite{stammer2024energy}. 
Furthermore, our work opens the door to investigate the interplay between the driving field symmetries and solid state systems, which naturally include translation, rotation, and reflection symmetries~\cite{ghimire_tutorial_2026}.

\begin{acknowledgments}

\emph{Acknowledgments.--}
ICFO-QOT group acknowledges support from:
MCIN/AEI (PGC2018-0910.13039/501100011033, CEX2019-000910-S/10.13039/501100011033, Plan National STAMEENA PID2022-139099NB, project funded MCIN and by the “European Union NextGenerationEU/PRTR" (PRTR-C17.I1), FPI); Ministry for Digital Transformation and of Civil Service of the Spanish Government through the QUANTUM ENIA project call - Quantum Spain project, and by the European Union through the Recovery, Transformation and Resilience Plan - NextGenerationEU within the framework of the Digital Spain 2026 Agenda; CEX2024-001490-S [MICIU/AEI/10.13039/501100011033]; Fundació Cellex;
Fundació Mir-Puig; Generalitat de Catalunya (European Social Fund FEDER and CERCA program; Barcelona Supercomputing Center MareNostrum (FI-2023-3-0024);
Funded by the European Union (HORIZON-CL4-2022-QUANTUM-02-SGA, PASQuanS2.1, 101113690, EU Horizon 2020 FET-OPEN OPTOlogic, Grant No 899794, QU-ATTO, 101168628), EU Horizon Europe Program (No 101080086 NeQSTGrant Agreement 101080086 — NeQST).

J~.R.-D.~acknowledges funding from UK Engineering and Physical Sciences Research Council (EPSRC) Funding, Grant UKRI2300 - Attosecond Photoelectron Imaging with Quantum Light.
\end{acknowledgments}

\bibliography{References.bib}{}

\appendix
\newpage 
\clearpage
\onecolumngrid

\begin{center}
    \textbf{SUPPLEMENTARY MATERIAL}
\end{center}

\section{Theoretical description}
In this work, we consider HHG driven by a bicircular $\omega-2\omega$ field, where the fundamental driving field and its second harmonic have opposite helicities and equal amplitude.~In contrast to semiclassical approaches, here we consider the case where one of the frequency components is prepared in a squeezed state along a linearly polarized direction.~Without loss of generality, we assume that squeezing is applied to the $\parallel$ component of the $2\omega$ mode, although the extension to squeezing of the $\omega$ mode or other combinations of $\{\omega,2\omega\}$ and $\{\parallel,\perp\}$ is straightforward.~The initial state of the driving field is then written as
\begin{equation}
	\ket{\Phi(t_0)}
		= \hat{D}_{\omega,L}(\alpha_\omega\sqrt{2})
				\otimes [\hat{D}_{2\omega,R}(\alpha_{2\omega}\sqrt{2})\hat{S}_{2\omega,\parallel}(\xi)]\ket{\bar{0}},
\end{equation}
where $\hat{D}_{\omega,\mu}(\alpha) = \exp[\alpha \hat{a}_{\omega,\mu}^\dagger - \alpha^*\hat{a}_{\omega,\mu}]$ and $\hat{S}_{\omega,\mu}(\xi)  = \exp[(\xi\hat{a}^2_{\omega,\mu} - \xi^*\hat{a}^{\dagger 2}_{\omega,\mu} )/2]$ are the displacement and squeezing operators acting on the optical mode $(\omega,\mu)$, respectively, both of which act on the multimode vacuum $\ket{\bar{0}}$.~To ensure that in the absence of squeezing the driving field exhibits the correct dynamical symmetry of a bicircular field, we impose $\abs{\alpha_{\omega}} = \abs{\alpha_{2\omega}}$.

Focusing on the $2\omega$ component and using that $\hat{D}_{2\omega,R}(\alpha_{2\omega}\sqrt{2}) \equiv \hat{D}_{2\omega,\parallel}(\alpha_{2\omega}) \otimes \hat{D}_{2\omega,\perp}(i \alpha_{2\omega})$, we insert the coherent state resolution of the identity $\mathbbm{1}_{\parallel} = \pi^{-1}\int \dd^2\alpha \dyad{\alpha}$ for the parallel polarization mode, such that the $2\omega$ component of the state becomes
\begin{equation}
	\ket{\Phi_{2\omega}(t_0)}
		= \frac{1}{\pi} \int \dd^2 \alpha \ c(\alpha) \hat{D}_{2\omega,\parallel}(\alpha)
            \otimes \hat{D}_{2\omega,\perp}(i\alpha_{2\omega})
        \ket{0_{2\omega,\parallel},0_{2\omega,\perp}},
\end{equation}
with $c(\alpha) = \bra{\alpha}\!\hat{D}_{2\omega,\parallel}( \alpha_{2\omega}) \hat{S}_{2\omega,\parallel}(r)\!\ket{0}$. The full initial state then reads
\begin{equation}
	\ket{\Phi(t_0)}
		= \dfrac{1}{\pi} 
		\int \dd^2 \alpha\ c(\alpha)
				\big[
					\hat{D}_{\omega,\parallel}(\alpha_{\omega}) \otimes \hat{D}_{\omega,\perp}(-i \alpha_{\omega})
				\big]
				\otimes
				\big[
					\hat{D}_{2\omega,\parallel}(\alpha) \otimes \hat{D}_{2\omega,\perp}(i\alpha_{2\omega})
				\big]
				\ket{\bar{0}}.
\end{equation}

The interaction with the atomic system is described in the length gauge, within the dipole and single-active-electron approximations, by the dipole coupling Hamiltonian
\begin{equation}
	\hat{H}(t)
		= \hat{H}_{\text{at}}
			+ \mathsf{e}\hat{\boldsymbol{r}}\cdot \hat{\boldsymbol{E}}(t),
\end{equation}
where $\hat{H}_{\text{at}}$ is the atomic Hamiltonian and $\hat{\boldsymbol{E}}(t) = - i \sum_{q,\mu}\kappa_q[\boldsymbol{\epsilon}_\mu \hat{a}_{q,\mu}e^{-i\omega_q t} + \text{h.c.}]$ the electric field operator, with $\boldsymbol{\epsilon}_\mu$ denoting the polarization vector and $\kappa_q \propto \sqrt{\omega_q/V}$ a constant depending on the quantization volume $V$.~This Hamiltonian is written in the interaction picture with respect to the free-field Hamiltonian $\hat{H}_{\text{field}} = \sum_{q,\mu}\hbar \omega_q \hat{a}^\dagger_{q,\mu}\hat{a}_{q,\mu}$.~The propagator $\hat{U}(t,t_0)$ time-evolving the state from the initial time $t_0$ up to the final time $t\geq t_0$ satisfies
\begin{equation}
	i\hbar \pdv{\hat{U}(t)}{t}
		= \hat{H}(t) \hat{U}(t),
\end{equation}
such that the evolved state of the joint light-matter system $\ket{\Psi(t)}$ reads
\begin{equation}\label{Eq:SM:evolved:I}
	\begin{aligned}
	\ket{\Psi(t)}
		&= \dfrac{1}{\pi}\int \dd^2 \alpha\ c(\alpha)
				\hat{U}(t,t_0)
				\big[
					\hat{D}_{\omega,\parallel}(\alpha_{\omega})
						 \otimes \hat{D}_{\omega,\perp}(-i \alpha_{\omega})
				\big]
				\otimes
				\big[
					\hat{D}_{2\omega,\parallel}(\alpha)
						 \otimes \hat{D}_{2\omega,\perp}(i\alpha_{2\omega})
				\big]
				\ket{\text{g}}\otimes\ket{\bar{0}}
		\\&
		= \dfrac{1}{\pi}\int \dd^2 \alpha\ c(\alpha)
				\big[
					\hat{D}_{\omega,\parallel}(\alpha_{\omega})
					\otimes \hat{D}_{\omega,\perp}(-i \alpha_{\omega})
				\big]
				\otimes
				\big[
					\hat{D}_{2\omega,\parallel}(\alpha)
					\otimes \hat{D}_{2\omega,\perp}(i\alpha_{2\omega})
				\big]
				\hat{U}(t,t_0;\alpha)
				\ket{\text{g}}\otimes\ket{\bar{0}}
		\\&
		= \dfrac{1}{\pi} \int \dd^2 \alpha\ c(\alpha)
				\hat{\boldsymbol{D}}(\boldsymbol{\alpha})
					\hat{U}(t,t_0;\alpha)
						\ket{\text{g}}\otimes \ket{\bar{0}},
	\end{aligned}
\end{equation}
where $\hat{\boldsymbol{D}}(\boldsymbol{\alpha})\equiv [\hat{D}_{\omega,\parallel}(\alpha_{\omega})\otimes \hat{D}_{\omega,\perp}(-i \alpha_{\omega})]\otimes
[\hat{D}_{2\omega,\parallel}(\alpha)
\otimes \hat{D}_{2\omega,\perp}(i\alpha_{2\omega})]$. Here, we have defined
\begin{equation}
	\begin{aligned}
	\hat{U}(t,t_0;\alpha)
		= \hat{\boldsymbol{D}}^\dagger(\boldsymbol{\alpha})
			\hat{U}(t,t_0)
			\hat{\boldsymbol{D}}(\boldsymbol{\alpha}),
	\end{aligned}
\end{equation}
with the transformed propagator satisfying
\begin{equation}
	i \hbar	\pdv{\hat{U}(t;\alpha)}{t}
		= 
		\Big[
			\hat{H}_{\text{at}} 
			+ \mathsf{e} \hat{\boldsymbol{r}}\cdot \hat{\boldsymbol{E}}(t)
			+ \mathsf{e} \hat{\boldsymbol{r}}\cdot \boldsymbol{E}_{\text{cl}}(t;\alpha)
		\Big]
		\hat{U}(t;\alpha),
\end{equation}
where $\boldsymbol{E}_{\text{cl}}(t;\alpha) = \bra{\bar{0}}\! \hat{\boldsymbol{D}}^\dagger(\boldsymbol{\alpha}) \hat{\boldsymbol{E}}(t)\hat{\boldsymbol{D}}(\boldsymbol{\alpha})\!\ket{\bar{0}}$ is the classical electric field from the quantum state defined by displacement the vacuum via $\hat{\boldsymbol{D}}(\boldsymbol{\alpha})$.

Restricting to HHG processes in which  the atom returns to the ground state, and working in the low-depletion regime such that time-dependent dipole correlations are negligible~\cite{stammer2024entanglement}, we obtain~\cite{rivera2022strong,stammer2023quantum}
\begin{equation}
	\bra{\text{g}}\!{\hat{U}(t,t_0;\alpha)}\!\ket{\text{g}}
		\simeq \bigotimes_q \hat{D}\big(\chi_q(t,t_0;\alpha)\big)
		\equiv \hat{\boldsymbol{D}}\big(\boldsymbol{\chi}_{q,\mu}(t,t_0)\big),
\end{equation}
where $\chi_{q,\mu}(t) \propto \int^{t}_{t_0}\dd \tau _q[\langle \hat{d}_{\mu}(\tau;\alpha)\rangle e^{-i\omega_q \tau}$.~Here, $\langle d_{\mu}(t;\alpha)\rangle = \langle \psi_{\alpha}(t)\vert \boldsymbol{\epsilon}_{\mu}\cdot\hat{\boldsymbol{d}}\vert \psi_\alpha(t)\rangle$ is the time-dependent dipole moment, with $\ket{\psi_{\alpha}(t)}$ the electronic state evolved through the semiclassical time-dependent Schrödinger equation (TDSE), i.e.,
\begin{equation}\label{Eq:SM:scl}
    i\hbar \pdv{\ket{\psi_\alpha(t)}}{t}
        = \big[
            \hat{H}_{\text{at}}
            + \mathsf{e} \hat{\boldsymbol{r}}\cdot \boldsymbol{E}_{\text{cl}}(t;\alpha)
        \big]
            \ket{\psi_\alpha(t)}.
\end{equation}
Therefore, the resulting quantum optical state $\ket{\Phi(t)} = \braket{\text{g}}{\Psi(t)}$ becomes
\begin{equation}
	\ket{\Phi(t)}
		= \dfrac{1}{\mathcal{N}\pi}
			\int \dd^2 \alpha \ c(\alpha)
				\hat{\boldsymbol{D}}(\boldsymbol{\alpha})
					\hat{\boldsymbol{D}}\big(\boldsymbol{\chi}(t,t_0;\alpha)\big)
						\ket{\bar{0}},
\end{equation}
where $\mathcal{N}^2 = \abs{\braket{\text{g}}{\Psi(t)}}^2$.~In the following, we shall denote for notational simplicity $\boldsymbol{\chi}_q(t,t_0;\alpha) = \boldsymbol{\chi}_q(\alpha)$. Using this notation, we find that the individual state of the $q$th harmonic order  obtained after tracing out all other optical modes reads
\begin{equation}
	\begin{aligned}
	\hat{\rho}_q(t)
		&= \dfrac{1}{\mathcal{N}^2\pi^2}
			\int \dd^2 \alpha \int \dd^2 \beta \ 
				c(\alpha)c^*(\beta)
					\bra{\bar{0}}\!
						\hat{\boldsymbol{D}}^\dagger(\boldsymbol{\chi}(\beta))
						\hat{\boldsymbol{D}}^\dagger(\boldsymbol{\beta})
						\hat{\boldsymbol{D}}(\boldsymbol{\alpha})
						\hat{\boldsymbol{D}}(\boldsymbol{\chi}(\alpha))
					\!\ket{\bar{0}}
					\bigotimes_{\mu}\dyad{\chi_{q,\mu}(\alpha)}{\chi_{q,\mu}(\beta)}
		\\&= \dfrac{1}{\mathcal{N}^2\pi^2}
					\int \dd^2 \alpha \int \dd^2 \beta \ 
						c(\alpha)c^*(\beta)
						w(\boldsymbol{\alpha},\boldsymbol{\beta},\boldsymbol{\chi})
						\bigotimes_{\mu}\dyad{\chi_{q,\mu}(\alpha)}{\chi_{q,\mu}(\beta)}
	\end{aligned} 
\end{equation}
where we have defined
\begin{equation}
	\begin{aligned}
	w(\boldsymbol{\alpha},\boldsymbol{\beta},\boldsymbol{\chi})
		&= 
		\bra{\bar{0}}\!
			\hat{\boldsymbol{D}}^\dagger\big(\boldsymbol{\chi}(\beta)\big)
            \hat{\boldsymbol{D}}^\dagger(\boldsymbol{\beta})
			\hat{\boldsymbol{D}}(\boldsymbol{\alpha})
			\hat{\boldsymbol{D}}\big(\boldsymbol{\chi}(\alpha)\big)
		\!\ket{\bar{0}}.
	\end{aligned}
\end{equation}

In general, the function above can be factorized into products of terms of the form
\begin{equation}\label{Eq:SM:field}
	\begin{aligned}
	\bra{0}
		\!\hat{D}^\dagger(\chi(\beta))\hat{D}^\dagger(\beta)
			\hat{D}(\alpha)\hat{D}(\chi(\alpha))\!
	\ket{0}
	&= e^{[\beta^*\chi(\beta)
				- \beta\chi^*(\beta)]/2}
		e^{[\alpha \chi^*(\alpha)
			- \alpha^*\chi(\alpha)]/2} \braket{\beta + \chi(\beta)}{\alpha + \chi(\alpha)}
	\\&
	= e^{-[\alpha-\beta]\chi^*(\beta) + [\alpha-\beta]^*\chi(\alpha)}
		\braket{\beta}{\alpha}
		\braket{\chi(\beta)}{\chi(\alpha)},
	\end{aligned}
\end{equation}
where, for harmonics $q > 2$, we set $\alpha = \beta = 0$ since these modes are initially not populated. From the explicit form of the coherent state overlap, we observe that the dominant contribution to $w(\boldsymbol{\alpha},\boldsymbol{\beta},\boldsymbol{\chi})$ arises from the region where $\alpha \approx \beta$, as the modulus of this function decays exponentially as $\exp[-|\beta-\alpha|^2/2]$. Already for variations $\abs{\beta-\alpha}  = 5$, the suppression is of the order of $10^{-6}$. In the present context, such variations correspond to changes in the driving field of only a few photons, or equivalently to electric field variations of $E_0 = 2 \kappa_q \alpha \sim 10^{-7}$ a.u.~By contrast, the strong-field response of the atom only becomes significantly modified for electric field changes of the order of $10^{-2}$ a.u., corresponding to variations of millions of photons.~Consequently, in the region where $w(\boldsymbol{\alpha},\boldsymbol{\beta},\boldsymbol{\chi})$ is appreciable, the harmonic amplitudes satisfy $\chi(\beta) \simeq \chi(\alpha)$.~Furthermore, in this regime the additional exponential factor contains terms proportional to $\abs{(\alpha-\beta)\chi^*(\alpha)} \propto \abs{\chi(\alpha)}$ and its complex conjugate. Since in the single-atom response considered here $\abs{\chi(\alpha)}\ll 1$, we may therefore approximate $e^{\abs{(\alpha-\beta)\chi^*(\alpha)}} \approx 1$.~Under these conditions, Eq.~\eqref{Eq:SM:field} simplifies to
\begin{equation}
	\bra{0}
		\!\hat{D}^\dagger(\chi(\boldsymbol{\beta}))\hat{D}^\dagger(\beta)
		\hat{D}(\alpha)\hat{D}(\chi(\boldsymbol{\alpha}))\!
	\ket{0}
	\simeq \braket{\beta}{\alpha}.
\end{equation}

Introducing this in Eq.~\eqref{Eq:SM:field}, we then arrive at
\begin{equation}\label{Eq:SM:harmonics:state}
	\begin{aligned}
	\hat{\rho}_{q,\mu}(t)
		&\simeq \dfrac{1}{\mathcal{N}^2\pi^2}
			\int \dd^2 \alpha
				\int \dd^2 \beta
					\ \braket{\alpha}{\Phi(t_0)}\braket{\Phi(t_0)}{\beta}\braket{\beta}{\alpha}
						\dyad{\chi_{q,\mu}(\alpha)}{\chi_{q,\mu}(\alpha)}
		\\&= \dfrac{1}{\pi}
			\int \dd^2 \alpha\
				\abs{\braket{\alpha}{\Phi(t_0)}}^2
					\dyad{\chi_{q,\mu}(\alpha)}{\chi_{q,\mu}(\alpha)}
		\\&=\int \dd^2 \alpha\
				Q(\alpha)
					\dyad{\chi_{q,\mu}(\alpha)}{\chi_{q,\mu}(\alpha)},
	\end{aligned}
\end{equation}
where $Q(\alpha) = \pi^{-1}\abs{\braket{\alpha}{\Phi(t_0)}}^2$ represents the Husimi function of the squeezed field mode in the driving field.

\section{Effect of classical selection rules on the electric field variance}
Following Eq.~\eqref{Eq:SM:harmonics:state}, we see that when considered individually and under the strong-field conditions required for their generation, the harmonics are effectively described by a classical state given by a statistical mixture of coherent states with amplitude $\chi_{q,\mu}(\alpha)$. The mean photon number of the emitted harmonics is therefore given by
\begin{equation}
	\langle \hat{a}^\dagger_{q,\mu}\hat{a}_{q,\mu}\rangle
		=\int \dd^2 \alpha\
			Q(\alpha)
				\abs{\chi_{q,\mu}(\alpha)}^2.
\end{equation}
From this expression we see that the harmonic radiation is determined by the average of the atomic response to classical fields $\boldsymbol{E}_{\text{cl}}(t;\alpha)$ sampled from the Husimi function $Q(\alpha)$. These atomic responses are ultimately governed by the semiclassical TDSE in Eq.~\eqref{Eq:SM:scl} which, for isotropic atomic systems, the presence of selection rules in the HHG response is determined by the dynamical symmetries obeyed by the driving field~\cite{neufeld2025light}.

Within the present framework, the conservation of selection rules in the presence of squeezed driving fields can be understood in terms of the extent to which the classical fields $\boldsymbol{E}_{\text{cl}}(t;\alpha)$ within the support of $Q(\alpha)$ satisfy the corresponding dynamical symmetries.~In other words, if the fluctuations of the field within $Q(\alpha)$ do not break the symmetry operations defining the dynamical symmetry group, the corresponding selection rules remain robust. Thus, to determine if a given non-classical driving field satisfies a certain dynamical symmetry, one must analyze whether the corresponding quantum state remains invariant under the relevant symmetry operations.~In practice, this can be assessed by studying quantities such as the expectation value of the field and its second-order moments, which characterize the mean field and its fluctuations.

For a bicircular field, the relevant dynamical symmetry operators are $\hat{r}_n$ and $\hat{\tau}_n$, which respectively describe a rotation of the electromagnetic field polarization by $2\pi/n$, and a time translation of $T/n$ with $T$ the fundamental period~\cite{neufeld2025light}.~Specifically, for bicircular fields, we have $n = 3$, meaning that simultaneously applying $\hat{r}_3$ and $\hat{\tau}_3$ leaves the field invariant.~These symmetries are responsible for the well-known selection rules in the HHG process driven by bicircular light. 

To extend these notions to our squeezed driving field, we need quantum optical versions of the above operators. For the polarization rotation, $\hat{r}_3$, we can define
\begin{equation}
	\hat{R}(\theta) 
        = \exp[-i \theta
                \big(
                    \hat{a}^\dagger_R\hat{a}_R
                    - \hat{a}^\dagger_L\hat{a}_L 
            \big)],
\end{equation}
where $\hat{a}_\mu$ denotes the annihilation operator acting on the polarization mode $\mu$, with $\mu=R (L)$ for the right-(left-) circularly polarized mode. For the time-translation operator, we have
\begin{equation}
	\hat{T}(\theta)
		= \exp[-i \theta \omega
					\big(
						\hat{a}^\dagger_R\hat{a}_R
						+ \hat{a}^\dagger_L\hat{a}_L 
					\big)],
\end{equation}
which corresponds to the propagation of the quantum state under the free-field Hamiltonian.

To evaluate the mean field and its fluctuations, we explicitly write the electric field operator of the driving field modes (highlight through the subscript ``d'') in the circularly polarized basis
\begin{equation}
	\hat{\boldsymbol{E}}_{\text{d}}(t)
		= \sum_{\omega}\sum_{\mu=R,L}
				\big[
					\boldsymbol{\epsilon}_\mu \hat{a}_{\omega,\mu} e^{-i\omega t}
					+ \boldsymbol{\epsilon}^*_\mu \hat{a}^\dagger_{\omega,\mu} e^{i\omega t}
				\big].
\end{equation}
With this definition, we can examine how the above dynamical symmetry transformations act on the operator above
\begin{equation}\label{Eq:EM:Field}
	\begin{aligned}
	\hat{\boldsymbol{E}}_{\text{d}}(t;\theta,\tau)
		&= \hat{T}^\dagger(\theta)\hat{R}^\dagger(\theta)
				\hat{\boldsymbol{E}}_{\text{d}}(t)
			\hat{T}(\theta)\hat{R}(\theta)
		\\&= \bigg[
				\sum_ \omega
				\boldsymbol{\epsilon}_R	\hat{a}_{R,\omega} e^{-i(\omega (t+\tau)+\theta)}
					+ \boldsymbol{\epsilon}_{R,\omega}^* \hat{a}^\dagger_Re^{i(\omega (t+\tau)+\theta)}
			\bigg]
			+
			\bigg[
				\sum_\omega
				\boldsymbol{\epsilon}_{L,\omega}	\hat{a}_L e^{-i(\omega (t+\tau)-\theta)}
				+ \boldsymbol{\epsilon}_{L,\omega}^* \hat{a}^\dagger_Le^{i(\omega (t+\tau)-\theta)}
			\bigg]
		\\&=\boldsymbol{\epsilon}_R
				\bigg[
					\sum_\omega 
					\hat{a}_{R,\omega} e^{-i(\omega (t+\tau)+\theta)}
					+ \hat{a}^\dagger_{L,\omega} e^{i(\omega(t+\tau)-\theta)}
				\bigg]
			+ \boldsymbol{\epsilon}_L
				\bigg[
					\sum_\omega
					\hat{a}_{L,\omega} e^{-i(\omega (t+\tau)-\theta)}
					+ \hat{a}^\dagger_{R,\omega} e^{i(\omega(t+\tau)-\theta)}
				\bigg],
	\end{aligned}
\end{equation}
where, in going from the second to the third equality, we have used that $\boldsymbol{\epsilon}_R = \boldsymbol{\epsilon}_L^*$.~Furthermore, we also need the variance of the electric field operator, which requires evaluating the expectation value of $\hat{\boldsymbol{E}}_{\text{d}}^2(t;\theta,\tau)$. From Eq.~\eqref{Eq:EM:Field}, this operator can be written as
\begin{equation}
	\begin{aligned}
		\hat{\boldsymbol{E}}_{\text{d}}^2(t;\theta,\tau)
			&= \bigg[
					\sum_\omega 
						\hat{a}_{R,\omega} e^{-i(\omega (t+\tau)+\theta)}
						+ \hat{a}^\dagger_{L,\omega} e^{i(\omega(t+\tau)-\theta)}
				\bigg]
				\bigg[
					\sum_\omega
						\hat{a}_{L,\omega} e^{-i(\omega (t+\tau)-\theta)}
						+ \hat{a}^\dagger_{R,\omega} e^{i(\omega(t+\tau)-\theta)}
				\bigg] + \text{h.c.}
			\\&
				= 2N_{\omega} + 
				2\bigg[
						\sum_\omega 
							\hat{a}_{R,\omega} e^{-i(\omega (t+\tau)+\theta)}
					\bigg]
					\bigg[
						\sum_\omega 
						\hat{a}_{L,\omega} e^{-i(\omega (t+\tau)-\theta)}
					\bigg]
				\\&
				\quad + 2 \bigg[
						\sum_\omega 
						\hat{a}^\dagger_{R,\omega} e^{i(\omega (t+\tau)+\theta)}
				\bigg]
					\bigg[
						\sum_\omega 
						\hat{a}_{R,\omega} e^{-i(\omega (t+\tau)+\theta)}
					\bigg]
				+2 
					\bigg[
						\sum_\omega 
						\hat{a}^\dagger_{L,\omega} e^{i(\omega (t+\tau)-\theta)}
					\bigg]
					\bigg[
						\sum_\omega 
							\hat{a}_{L,\omega} e^{-i(\omega (t+\tau)-\theta)}
					\bigg]
				\\&\quad +
					 2\bigg[
						 \sum_\omega 
					 		\hat{a}^\dagger_{R,\omega} e^{i(\omega (t+\tau)+\theta)}
					 \bigg]
					 \bigg[
				 		\sum_\omega 
					 		\hat{a}^\dagger_{L,\omega} e^{i(\omega (t+\tau)-\theta)}
					 \bigg],
	\end{aligned}
\end{equation} 
where $N_{\omega}$ denotes the total number of frequencies considered.~In going from the first to the second equality, we have used that $[\hat{a}_{\mu,\omega},\hat{a}^\dagger_{\mu,\omega}] = 1$.

In our configuration, we drive the HHG process with the state given in Eq.~\eqref{Eq:SM:field}, that is, a ``standard'' bicircular field in which some amount of squeezing is applied along a linearly polarized component (here $\parallel$, although it is irrelevant which linear direction is chosen).~Before proceeding, it is worth recalling the following transformations
\begin{equation}
	\hat{D}(\alpha)\hat{E}(t) \hat{D}(\alpha)
		= \hat{E}(t) + \bra{\alpha}\!\hat{E}(t)\!\ket{\alpha},
	\quad
	\hat{S}^\dagger(r) \hat{a}\hat{S}(r)
		= \hat{a} \cosh(r) - \hat{a}^\dagger \sinh(r),
\end{equation}
that is, the applied displacement operations recover a classical bicircular field, $\boldsymbol{E}_{\text{bic}}(t) \equiv \bra{\alpha_{\omega,R},\alpha_{2\omega,L}}\!\hat{\boldsymbol{E}}_{\text{d}}(t)\!\ket{\alpha_{\omega,R},\alpha_{2\omega,L}}$. With this in mind, one finds
\begin{equation}
	\langle\hat{\boldsymbol{E}}_{\text{d}}(t;\theta,\tau)\rangle
		= \boldsymbol{E}_{\text{bic}}(t;\theta,\tau)
			+ \bra{0}\!\hat{S}^\dagger(r)\hat{E}(t;\theta,\tau)\hat{S}(r)\!\ket{0}
		= \boldsymbol{E}_{\text{bic}}(t;\theta,\tau),
\end{equation}
so that for the classical bicircular component
\begin{equation}
	\boldsymbol{E}_{\text{bic}}(t;\theta,\tau)
		= \boldsymbol{\epsilon}_R\alpha
			\big[
				e^{-i(\omega(t+\tau)+\theta)}
				+ e^{i(2\omega(t+\tau)-\theta)}
			\big]
			+ \text{c.c.},
\end{equation}
from which we can see that by setting $\theta=2\pi/3$ and $\tau = 2\pi/(3\omega)$ the field remains invariant.~More explicitly, we find that
\begin{equation}
	\boldsymbol{E}_{\text{bic}}\big(t;2\pi/3,2\pi/(3\omega)\big)
		= e^{i2\pi/3}\alpha \boldsymbol{\epsilon}_R
			\big[
				e^{-i\omega t}
				+ e^{i2\omega t}
			\big]
			+ \text{c.c.}
		\equiv \boldsymbol{E}_{\text{bic}}(t),
\end{equation}
which is $2\pi/3$ rotation of the polarization axes, leaving the time dependence unchanged.~Hence, on average, the field satisfies the dynamical symmetry.

\begin{figure}
    \centering
    \includegraphics[width=1\textwidth]{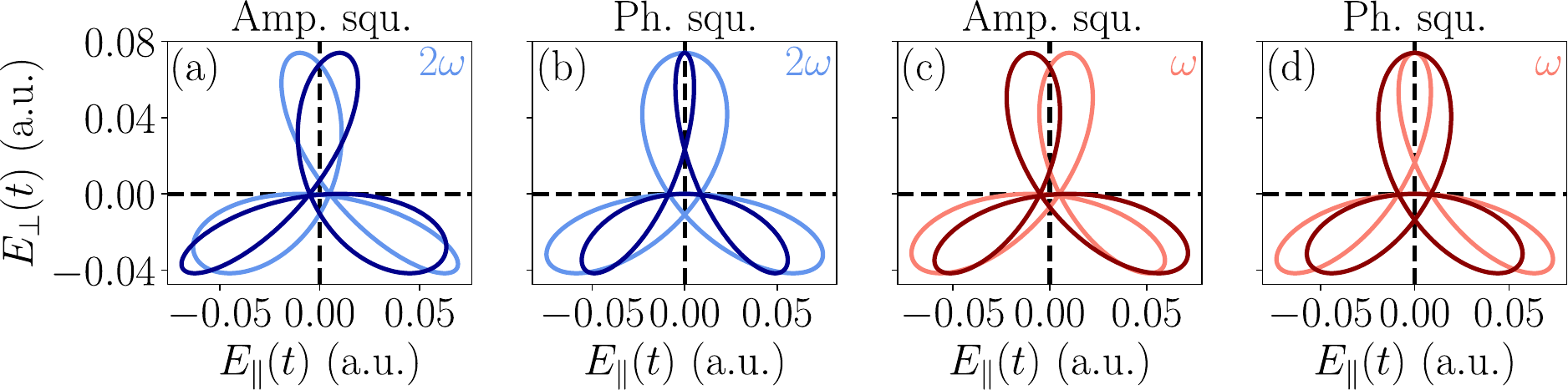}
    \caption{Lissajous figures for two values of the electric field sampled symmetrically around the maxima of $Q(\alpha)$. Each panel shows a different driving field configuration: (a) and (b) correspond to the case where amplitude and phase squeezing are applied to the $2\omega$ field, respectively, while panels (c) and (d) show the case where it is applied to the $\omega$ field instead.}
    \label{Fig:Lissajous:Extremes}
\end{figure}

However, the same cannot be said for the fluctuations.~For the expectation values of the squared electric field operator, we find
\begin{equation}
	\begin{aligned}
	\langle
		\hat{\boldsymbol{E}}^2(t;\theta,\tau)
	\rangle
		&= \boldsymbol{E}^2_{\text{bic}}(t)
			+ 2 \boldsymbol{E}_{\text{bic}}(t)
				\bra{0}\!\hat{S}^\dagger(r)\hat{E}(t;\theta,\tau)\hat{S}(r)\!\ket{0}
			+  \bra{0}\!\hat{S}^\dagger(r)\hat{\boldsymbol{E}}^2(t;\theta,\tau)\hat{S}(r)\!\ket{0}
		\\&
			= \boldsymbol{E}^2_{\text{bic}}(t)
				+ \bra{0}\!\hat{S}^\dagger(r)\hat{\boldsymbol{E}}^2(t;\theta,\tau)\hat{S}(r)\!\ket{0}.
	\end{aligned}
\end{equation}
Using $\hat{a}_{R/L} = (\hat{a}_\parallel \mp i \hat{a}_\perp)/\sqrt{2}$, we obtain
\begin{equation}
	\bra{0}\!\hat{S}^\dagger(r)\hat{\boldsymbol{E}}^2(t;\theta,\tau)\hat{S}(r)\!\ket{0}
		= 4\big\{
				1 + \sinh(r)
					\big[\sinh(r) - \cosh(r)\cos(4\omega(t+\tau))\big]
			 \big\},
\end{equation}
which gives a field variance
\begin{equation}\label{Eq:SM:fluctuations}
	\Delta E^2(t;\theta,\tau)
		= \langle \hat{\boldsymbol{E}}^2(t;\tau,\theta)\rangle
			-  \langle \hat{\boldsymbol{E}}(t;\tau,\theta)\rangle^2
		= 4\Big\{
				1 + \sinh(r)
					\big[\sinh(r) - \cosh(r)\cos(4\omega(t+\tau))\big]
				\!
			\Big\},
\end{equation}
and that, as we can see, does not respect the dynamical symmetry satisfied by the mean field.~Thus, while the average field satisfies the dynamical symmetry, the fluctuations of the field do not.~Consequently, when the field fluctuations are large, the selection rules expected for HHG under a classical bicircular driver are no longer satisfied.~These fluctuations play a crucial role in determining the HHG response, allowing harmonics that would otherwise be forbidden by the classical symmetry.~As representative examples and using the Husimi function representation, Fig.~\ref{Fig:Lissajous:Extremes} presents the type of Lissajous figures that would be obtain for electric fields located symmetrically around the maxima of $Q(\alpha)$, in light and dark colors, for different field configurations. 

Having justified the fluctuation induced breaking of the dynamical symmetries of the driving field, we now discuss how this affects the harmonic generation process.~From Eq.~\eqref{Eq:SM:fluctuations}, we observe that the fluctuations are invariant under the joint operation of a temporal translation $\hat{\tau}_4 = \pi/(4\omega)$.~Taking into account the full field properties by means of the classical average, being invariant under $\hat{r}_3\hat{\tau}_3$, and the field fluctuations, with invariance under $\hat{\tau}_4$, the intersection of the two symmetry groups reduces to the identity.   
As a result, no nontrivial dynamical symmetry is given for full driving field, implying that the HHG process is only periodic over a full optical cycle. Consequently, all harmonic orders of $\omega$ are, in principle, allowed, as shown in the main text. 

The above can be seen more explicitly by considering Eq.~\eqref{Eq:SM:harmonics:state}, which also reflects how calculation is performed in practice. This state corresponds to a classical mixture of coherent states weighted by the Husimi function, which determines the driving field used to evaluate the harmonic spectral amplitude. More specifically, the Husimi function samples driving fields of the form
\begin{equation}\label{Eq:SM:Husimi:fields}
    \boldsymbol{E}_{\text{cl}}(t;\alpha) = 
        \underbrace{\boldsymbol{E}_{\text{bic}}(t;\alpha)}_{\tau_3 r_3}
        + \underbrace{\boldsymbol{E}_{2\omega,\parallel}(t)}_{\tau_2 r_2},
\end{equation}
that is, a linearly polarized component added to a bicircular field, with its strength dictated by the Husimi distribution. As discussed above, the only remaining symmetry of this field is the trivial temporal over a full optical cycle. In the absence of the linearly polarized component, the spectral intensity for a bicircular field can be written as
\begin{equation}
    \chi_{q,\mu}
        \propto \sum_{n} 
            \int^{2\pi/\omega}_0 \dd t
                D_\mu(t) e^{-i q\omega t}
        = \sum_{n}
            \bigg[
                \int^{\frac{2\pi}{3\omega}}_{0} \dd t
                    D_{\mu}(t) e^{-iq\omega t}
                + \int^{\frac{4\pi}{3\omega}}_{\frac{2\pi}{3\omega}} \dd t
                    D_{\mu}(t) e^{-iq\omega t}
                + \int^{\frac{2\pi}{\omega}}_{\frac{4\pi}{3\omega}} \dd t
                    D_{\mu}(t) e^{-iq\omega t}
            \bigg].
\end{equation}
Assuming identical response across cycles, we split the integral into three intervals of duration $2\pi/(3\omega)$, yielding
\begin{equation}\label{Eq:SM:bicircular:spec}
     \chi_{q,\mu}
        \propto \sum_{n}
            \bigg[
                \int^{\frac{2\pi}{3\omega}}_{0} \dd t
                    D_{\mu}(t) e^{-iq\omega t}
                + \int^{\frac{2\pi}{3\omega}}_{0}\dd t
                    D_{\mu}\big(t + 2\pi/(3\omega)\big) e^{-iq\omega t - iq\frac{2\pi}{3}}
                + \int^{\frac{2\pi}{3\omega}}_{0} \dd t
                    D_{\mu}\big(t + 4\pi/(3\omega)\big) e^{-iq\omega t-iq\frac{4\pi}{3}}
            \bigg].
\end{equation}
Then, using the symmetry of the bicircular field
\begin{equation}
    D_{\mu}\big(t + 2\pi/(3\omega)\big) 
        = e^{\pm i \frac{2\pi}{3}}
            D_{\mu}(t)
    \quad 
        \text{and}
    \quad
    D_{\mu}\big(t + 4\pi/(3\omega)\big) 
        = e^{\pm i \frac{4\pi}{3}}
            D_{\mu}(t),
\end{equation}
where the $\pm$ depends on whether we are considering the left or circular polarization components, we therefore arrive at
\begin{equation}
    \chi_{q,\mu}
        \propto \sum_{n}
                \int^{\frac{2\pi}{3\omega}}_{0} \dd t
                    D_{\mu}(t) e^{-iq\omega t}
                        \Big[
                            1 + e^{\pm iq\frac{2\pi}{3}} + e^{\pm iq\frac{4\pi}{3}}
                        \Big],
\end{equation}
where the term in brackets is maximized whenever $q = 3 N \pm 1$, with $N\in \mathbbm{W}$, while it vanishes when $q = 3N$, recovering the standard bicircular selection rules.

When a non-vanishing linearly polarized $2\omega$ component is included, the harmonic response must be evaluated using fields of the form in Eq.~\eqref{Eq:SM:Husimi:fields}, with amplitudes sampled from the Husimi function.~For clarity, assuming $\abs{\boldsymbol{E}_{\text{bic}}} \gg \abs{\boldsymbol{E}_{2\omega,\parallel}}$, we treat the additional linear component perturbatively, such that the harmonic amplitude can be written as~\cite{dahlstrom_quantum_2011}
\begin{equation}
     \chi_{q,\mu}
        \propto \sum_{n} 
            \int^{2\pi/\omega}_0 \dd t
                D_\mu(t) e^{-i \sigma(t) -i q\omega t},
\end{equation}
where $D_{\mu}(t)$ represents the bicircular contribution, while $\sigma(t)$ is a phase induced by the weak linear component. Repeating the previous decomposition, we obtain
\begin{equation}
    \chi_{q,\mu}
        \propto\sum_{n}
                \int^{\frac{2\pi}{3\omega}}_{0} \dd t
                    D_{\mu}(t) e^{-iq\omega t}
                        \Big[
                            e^{-i\sigma(t)} 
                            + e^{-i\sigma(t+\frac{2\pi}{3\omega})
                                \pm iq\frac{2\pi}{3}}
                            + e^{-i\sigma(t+\frac{4\pi}{3\omega})
                                \pm iq\frac{4\pi}{3}}
                        \Big].
\end{equation}
In contrast to the bicircular case, the additional phase $\sigma(t)$ breaks the threefold symmetry, preventing the exact cancellation that enforces the presence of specific harmonic orders. As a result, harmonics with $q=3N$, which are forbidden in the perfectly symmetric case, become allowed, although their intensity remains controlled by the strength of the symmetry-breaking perturbation.

\begin{figure}
    \centering
    \includegraphics[width=1\textwidth]{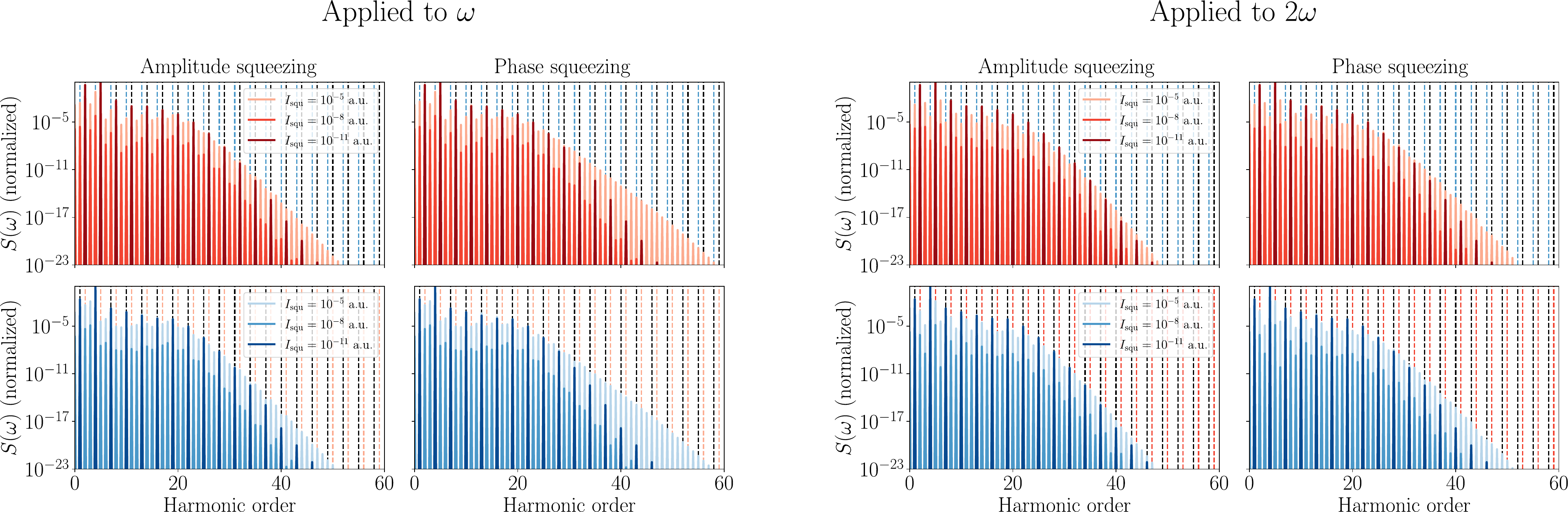}
    \caption{HHG spectra obtained when applying the linearly polarized squeezing to the $\omega$ (left panels) and to the $2\omega$ (right panels) field contribution. The first row displays the spectra evaluated along the RCP component, while the second row when evaluated along the LCP component.~The black dashed vertical lines highlight the locatiton of the classically allowed harmonics.~The atomic medium is hydrogen ($I_p = 0.5$ a.u.), the driving frequency is $\omega_L = 0.057$ a.u., and the bicircular component field amplitudes are set to $E_0 = 0.037$ a.u.}
    \label{Fig:SM:All:spectra}
\end{figure}

We emphasize that these results are independent of the specific mode to which the linearly polarized squeezing is applied, as well as of its relative phase with respect to the coherent state component.~This is explicitly illustrated in Fig.~\ref{Fig:SM:All:spectra}, which shows the HHG spectra obtained for both phase and amplitude squeezing, applied either to the $\omega$ field (left panels) or to the $2\omega$ component (right panels).~Only minor differences are observed, primarily in the cutoff regions, where phase squeezing leads to slightly more intense harmonic orders beyond the cutoff.

\section{Selection rules via spin angular momentum conservation}
The presence of specific selection rules in the harmonic generation process can also be understood from an interplay between the allowed harmonic frequencies and the conservation of specific quantities, such as is spin angular momentum (SAM) in our case~\cite{fleischer_spin_2014,pisanty_spin_2014}.~This perspective often provides a simple picture for understanding the polarization properties of the emitted harmonics.

For the case of a bicircular driver, composed of two circularly polarized components with different frequency, the harmonic generation process can be interpreted as the absorption (or stimulated emission) of photons from each field component while satisfying SAM conservation.~Denoting the $n_1$ and $n_2$ the number of photons absorbed (if $n_i > 0$) or stimulatedly emitted (if $n_i < 0$) from each channel, the generated harmonic satisfies
\begin{equation}
	\Omega = (n_1 + 2n_2)\omega, \quad \sigma = n_1 \sigma_1 + n_2 \sigma_2,
\end{equation}
where $\Omega$ is the harmonic frequency $\sigma$ its SAM.~In our case, we have that $\sigma_1 = -1$ (LCP) and $\sigma_2 = 1$ (RCP), and enforcing SAM conservation then leads to the well-known selection rule
\begin{equation}
	\Omega = (3n_1 \pm 2)\omega
\end{equation}
with the emitted harmonic carrying helicity $\sigma = \pm 1$. All other intermediate harmonics $\Omega = 3 n$ $\forall n\in \mathbbm{W}$ are forbidden.

Following from Eq.~\eqref{Eq:SM:harmonics:state}, when squeezing is applied along a linearly polarized direction, the quantum optical state of the harmonics is given by an incoherent mixture in which the harmonic generation process driven by the field $\boldsymbol{E}_{\text{bic}}(t) + \boldsymbol{\epsilon}_{\parallel}\abs{\varepsilon}\cos(2\omega t + \phi)$ occurs with probability $Q(\varepsilon/(2\kappa))$.~Since $\varepsilon_{\parallel} = (\varepsilon_{R} + \varepsilon_{L})/\sqrt{2}$, the total field can be expanded in the circularly polarized basis as
\begin{equation}
	\boldsymbol{E}_{\text{tot}}
		= 
		\big[
			E_{\omega}(t)+\delta E_{2\omega}(t) 
		\big]\boldsymbol{\epsilon}_L
		+ 
		\big[
			E_{2\omega}(t)
			+ \delta E_{2\omega}(t)
		\big]\boldsymbol{\epsilon}_R,
\end{equation}
where $\delta E_{2\omega}(t) = \abs{\varepsilon_{\parallel}}\cos(2\omega t + \phi)/\sqrt{2}$. We therefore see that the effect of squeezing is to introduce additional $2\omega$ components with both helicities.~As a consequence, new photon exchange pathways become available compared to the classical bicircular configuration, since photons at frequency $2\omega$ can now be absorbed (or stimulatedly emitted) with either polarization handedness~\cite{pisanty_spin_2014}.~Denoting by $n_{2,+}$ and $n_{2,-}$ the number of photons exchanged from the right- and left-handed $2\omega$ components, respectively, the harmonic generation process satisfies
\begin{equation}
	\Omega = (2 n_{2,+} + 2n_{2,-} + n_1)\omega,
		\quad
	\sigma = n_{2,+} - n_{2,-}-n_{1},
\end{equation}
which leads to
\begin{equation}\label{Eq:SM:channel:2w}
	\Omega = (3 n_1 + 4 n_{2,-} \pm 2)\omega,
		\quad
	n_{2,+} = \pm1 + n_1 + n_{2,-}.
\end{equation}

Thus, depending on the harmonic mode, several photon-exchange channels can contribute to the generation process. To understand what determines the helicity of the harmonics that are absent in the classical bicircular case, i.e., the $\{3 q\omega; q\in \mathbbm{N}\}$, we restrict our analysis to perturbative channels whose intensity scales as $I_{\Omega} \sim E_{\omega}^{2\abs{n_1}} \delta E_{2\omega,\parallel}^{2\abs{n_{2,-}}}  (E_{2\omega,R} + \delta E_{2\omega,\parallel})^{2\abs{n_{2,+}}}$.~In the regime $\delta E_{2\omega,\parallel} \ll E_{2 \omega}$, the dominant channels are those that minimize the power of the weak field component, i.e., those with the smallest possible value of $n_{2,-}$ or, when this is not possible, those that maximize $n_{1}$ and $n_{2,+}$. 

\begin{figure}[h!]
    \centering
    \includegraphics[width=1\textwidth]{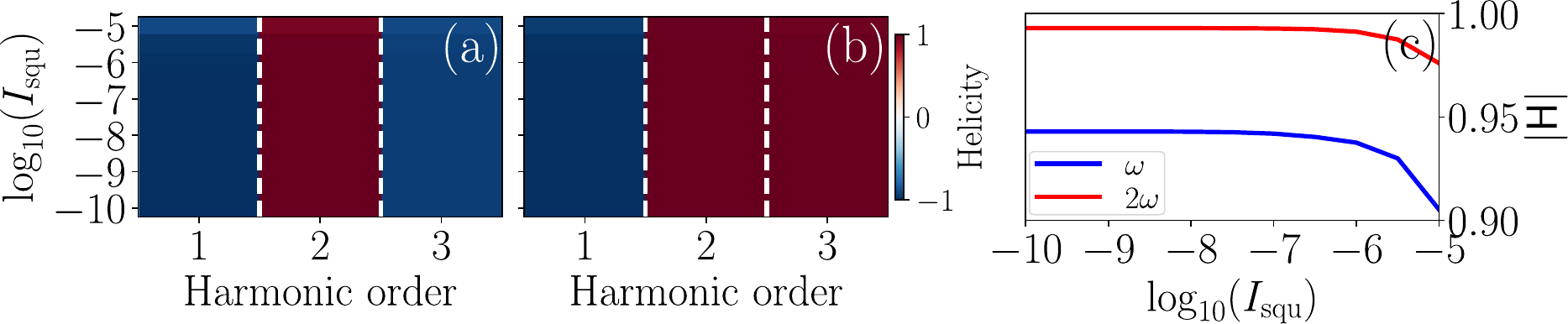}
    \caption{(a),(b) Helicity of the first three harmonic orders when driven introducing linearly polarized phase squeezing along the (a) $\omega$ and (b) $2\omega$ field components. (c) Helicity of the generated 3rd harmonic order as a function of the squeezing intensity.}
    \label{Fig:SM:Helicity:SAM}
\end{figure}

Considering the lowest values of $n_{2,-}$ compatible with Eq.~\eqref{Eq:SM:channel:2w}, we identify two possible channels
\begin{equation}
	\begin{aligned}
		&\text{\ding{172}}\  \sigma = +1 \implies n_1 = -1,\ n_{2,-} = 1, \ n_{2,+} = 1,\\
		& \text{\ding{173}}\ \sigma = -1 \implies n_1 = -1, \ n_{2,-} = 2, \ n_{2,+} = 0.
	\end{aligned}
\end{equation}
We therefore find that the most probable channel for generating the $3\omega$ photon is \ding{172}, resulting in a RCP photon.~This behavior is confirmed in Fig.~\ref{Fig:SM:Helicity:SAM}~(b), which shows the helicity of the first three harmonics as a function of the squeezing intensity.~As the squeezing increases, larger values of $\delta E_{2\omega,\parallel}$ are sampled, enhancing the contribution of channel \ding{173} and therefore reducing the helicity [Fig.~\ref{Fig:SM:Helicity:SAM}~(c) red curve].

To confirm this behavior, we consider the case where squeezing is applied onto the $\omega$ component instead.~In this scenario, the generation of a harmonic $\Omega$ satisfies
\begin{equation}\label{Eq:SM:channel:w}
	\Omega = (3 n_{1,-} - n_{1,+} \pm 2)\omega,
		\quad
	n_{2} = \pm1 + n_{1,-} - n_{1,+},
\end{equation}
with the intensity of the perturbative harmonics scaling as $I_{\Omega} \sim [E_{\omega} + \delta E_{\omega,\parallel}]^{2\abs{n_{1,-}}} \delta E_{\omega,\parallel}^{2\abs{n_{1,+}}}  E_{2\omega}^{2\abs{n_{2}}}$. The most probable channels are those that minimize $n_{1,+}$ or, if not possible, maximize the other two contributions.~Considering the lowest value of $\abs{n_{1,+}}$ we identify two possible channels
\begin{equation}
	\begin{aligned}
		&\text{\ding{172}}\  \sigma = +1 \implies \{n_{1,-} = 1,\ n_{1,+} = 2, \ n_{2} = 0\},\\
		& \text{\ding{173}}\ \sigma = -1 \implies \{n_{1,-} = -1, \ n_{1,+} = -2, \ n_{2,+} = 2\}.
	\end{aligned}
\end{equation}
In this case, both channels are degenerate in $\abs{n_{1,+}}$.~However, \ding{172} does not absorb photons from the strong $2\omega$ field whereas \ding{173} does.~Consequently, the most likely channel is \ding{173} which results in a LCP photon, as confirmed numerically in Fig.~\ref{Fig:SM:Helicity:SAM}~(b). As the amount of squeezing increases, the contribution of \ding{172} becomes more significant, thereby reducing the net helicity [Fig.~\ref{Fig:SM:Helicity:SAM}~(c) blue curve].

\begin{figure}[h!]
    \centering
    \includegraphics[width=1\textwidth]{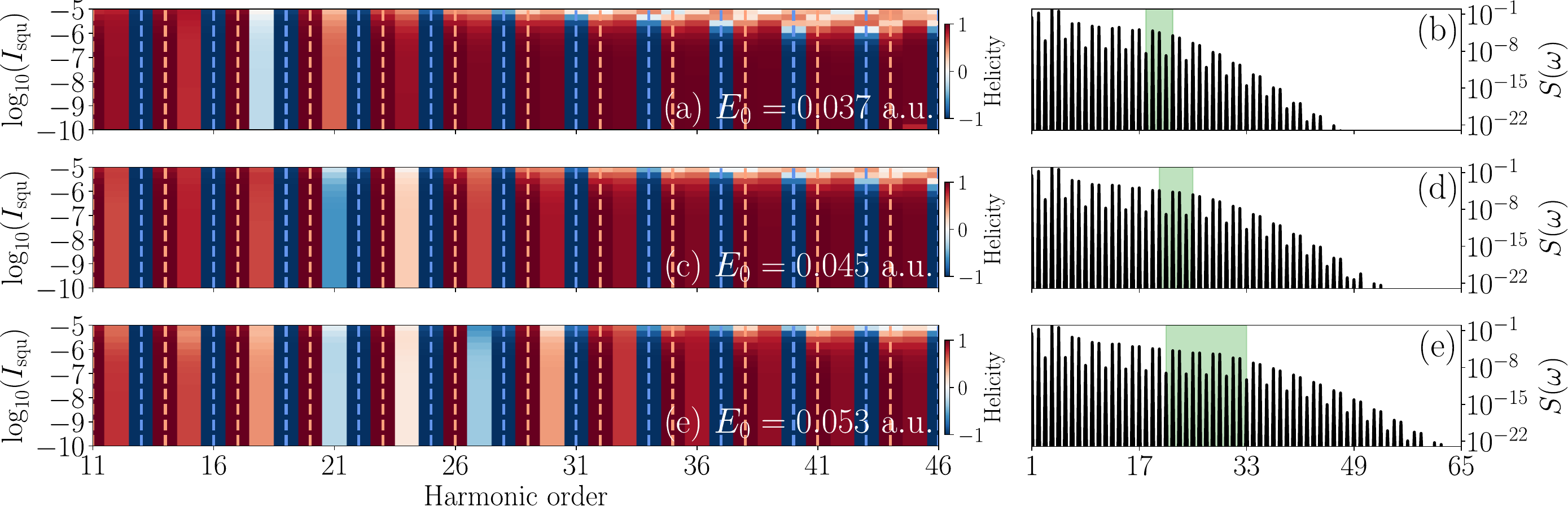}
    \caption{Panels (a), (c), (e) show the helicity for several high harmonic orders as a function of the squeezing intensity, with each panel corresponding to a different total intensity of the driving field.~Panels (b), (d) (e) display the corresponding HHG spectra obtained for $I_{\text{squ}} = 10^{-8}$ a.u., with the green region highlights those harmonics for which the helicity of the $3q$ ($q\in \mathbbm{N}$ harmonics deviates most strongly from $\mathsf{H} = 1$.}
    \label{Fig:SM:Helicity:for:fields}
\end{figure}

While this discussion provides a rationale for the preferred helicity of the generated harmonics, it is important to emphasize that the exact helicity depends on how the harmonic intensity scales with the various components of the driving field. In the plateau region, this scaling does not necessarily follow the perturbative form considered here. Indeed, we find that the helicity is particularly sensitive for harmonics located in the cutoff region, as illustrated in Fig.~\ref{Fig:SM:Helicity:for:fields}, which displays the helicity of plateau and cutoff harmonics for different bicircular field strengths.

\section{Photon statistics of the emitted harmonics}
As shown in the main text, for the bicircular configuration studied here it appears that the fluctuation induced harmonic modes appear to follow the same type of photon statistics as the driving field.~This occurs subject to an important difference: the emitted harmonics lie in classical states given by statistical mixtures of various coherent states. In this context, we can express the zero-delay second-order autocorrelation function $g_{q,\mu}^{(2)}(0)$ for a given harmonic mode $(q,\mu)$ as
\begin{equation}\label{Eq:SM:g2}
	g_{q,\mu}^{(2)}(0)
		= \dfrac{\langle \hat{a}_{q,\mu}^{\dagger 2}\hat{a}_{q,\mu}^2\rangle}{\langle \hat{a}_{q,\mu}\hat{a}_{q,\mu}\rangle^2}
		= \dfrac{\int \dd^2 \alpha\ Q(\alpha) \abs{\chi_{q,\mu}(\alpha)}^4}{\big[\int \dd^2 \alpha \ Q(\alpha)  \abs{\chi_{q,\mu}(\alpha)}^2\big]^2}.
\end{equation}

\begin{figure}
	\centering
	\includegraphics[width=1\columnwidth]{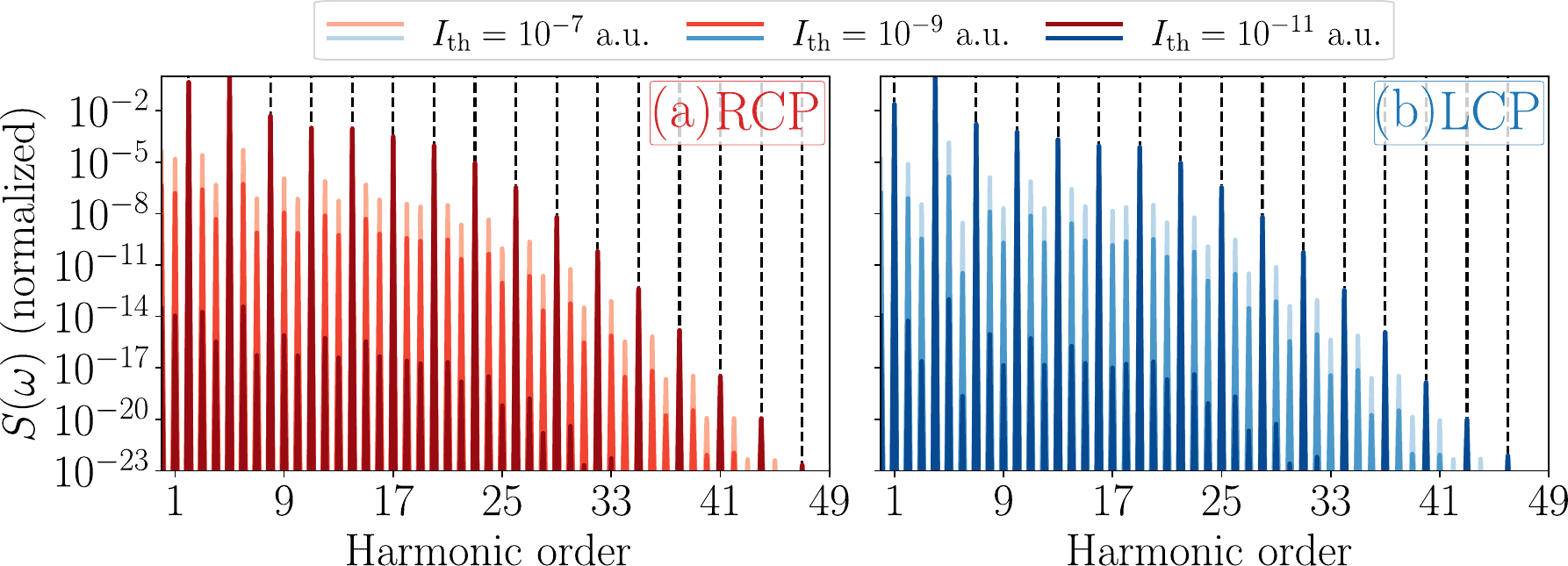}
	\caption{(a), (b) HHG spectra resolved along the $R$- and $L$-polarization components, respectively, when considering a thermal state as a driving field.~The atomic medium is hydrogen ($I_p = 0.5$ a.u.), the driving frequency is $\omega_L = 0.057$ a.u., and the bicircular coherent field amplitudes are set to $E_0 = 0.037$ a.u.}
	\label{Fig:SM:thermal}
\end{figure}

In the main text, we distinguish between two cases: one in which the bicircular displacement acts on an initially linearly polarized BSV state, and another in which it acts on a thermal state (with harmonics shown in Fig.~\ref{Fig:SM:thermal}).~For these two situations, the corresponding Husimi function can be generally written
\begin{equation}
	Q(\alpha)
		= \dfrac{1}{2\pi \sqrt{\sigma_x\sigma_y}}
				\exp[- \dfrac{\alpha_x^2}{2\sigma_x} - \dfrac{\alpha^2_y}{2\sigma_y}],
\end{equation}
where $\alpha = \alpha_x + i \alpha_y$.~For a thermal state we have that $\sigma_x = \sigma_y$, i.e., field fluctuations are equally distributed along all phase-space directions.~In contrast, for a squeezed state we have $\sigma_x > \sigma_y$ or vice versa, depending on which optical quadrature is squeezed. 

Without loss of generality, let us take the squeezed quadrature to correspond to $\sigma_y$.~In our case the applied squeezing is extremely large, such that $\sigma_y \ll 1$ (and, due to the Heisenberg uncertainty principle, $\sigma_x \gg1$).~Consequently, the Husimi function becomes extremely localized along the squeezed quadrature, such that the factor $\exp[-\alpha_y^2/(2\sigma_y)]$ rapidly decays to zero in a region where $\chi(\alpha)$ remains essentially unperturbed.~As a result, for strongly squeezed states we can approximate $\chi_{q,\mu}(\alpha)\simeq \chi_{q,\mu}(\alpha_x)$, which allows us to rewrite Eq.~\eqref{Eq:SM:g2} as
\begin{equation}\label{Eq:SM:g2:squ}
	g^{(2,\text{squ})}_{q,\mu}(0)
		= \dfrac{\int \dd \alpha_x\ \mathcal{Q}(\alpha_x) \abs{\chi_{q,\mu}(\alpha_x)}^4}{\big[\int \dd^2 \alpha \ \mathcal{Q}(\alpha_x)  \abs{\chi_{q,\mu}(\alpha_x)}^2\big]^2},
\end{equation}
where $\mathcal{Q}(\alpha_x)$ denotes the marginal of the Husimi function.~For a thermal state, however, such an approximation is not valid and the full 2D integral must be retained.

\begin{figure}
    \centering
    \includegraphics[width=1\textwidth]{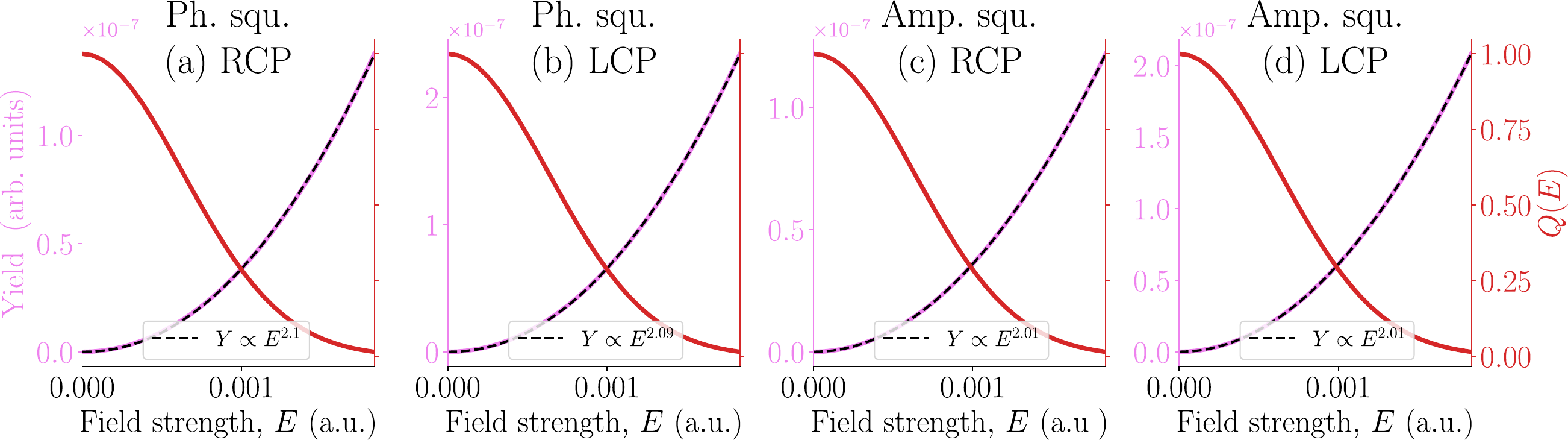}
    \caption{Scaling of the harmonic intensity of the 18th harmonic mode, which in the absence of squeezing would not appear under the bicircular configuration, as a function of the field strength. The exact scaling is shown in violet, while the fitting to $Y = A E^n$ is shown with the dashed black line. In red, the marginal of the Husimi function.}
    \label{Fig:SM:Yield}
\end{figure}

Focusing first on the case of squeezed states, let us assume that the harmonic response of the fluctuation induced harmonics has the form $\chi_{q,\mu}(\alpha) = A \alpha_x^n$. It then follows from Eq.~\eqref{Eq:SM:g2:squ} that~\cite{petrovic_generation_2026},
\begin{equation}
	g^{(2,\text{squ})}_{q,\mu}(0)
		= \sqrt{\pi}\dfrac{\Gamma(\tfrac12 + 2n)}{\Gamma(\tfrac12 + n)^2},
\end{equation} 
valid for $n\in \mathbbm{N}$.~From this expression, we find that when $n = 1$, $g^{(\text{squ})}_{q,\mu}(0) =3$, which coincides with the photon statistics expected for squeezed vacuum states of extremely large mean photon number.~This scaling is indeed confirmed numerically, as shown for the 18th harmonic in Fig.~\ref{Fig:SM:Yield}, where the harmonic intensity is plotted as a function of the electric field strength.~Therefore, it is the initial linear scaling of $\chi_{q,\mu}(\alpha)$ with respect to $\alpha$ when moving away from the exact bicircular configuration that enables the harmonic modes to inherit the photon statistics of the driving field, despite the fact that their quantum states are intrinsically different.~We note, however, that for field strengths larger than those shown in Fig.~\ref{Fig:SM:Yield} this linear scaling no longer holds.~This explains why $g^{(\text{squ})}_{q,\mu}(0)\neq 3$ in general for extremely large squeezing values. In that regime, the Husimi function includes field configurations that strongly break the initial bicircular-induced symmetry of the field.

It is worth highlighting that the scaling found in Fig.~\ref{Fig:SM:Yield} is independent on whether phase or amplitude squeezing is used, since both cases break the initial symmetry of the bicircular field.~We may therefore generalize the scaling to $\chi_{q,\mu}(\alpha) = A(\alpha_x^n + \alpha_y^n)$, such that for thermal fields, Eq.~\eqref{Eq:SM:g2} yields
\begin{equation}
	g^{(2,\text{th})}_{q,\mu}(0)
		= \dfrac{\Gamma(\tfrac12 + n)^2 + \sqrt{\pi}\Gamma(\tfrac12 + 2 n)}{2\Gamma(\tfrac12 + n)^2}
        = \dfrac12
            \big[
                1 + g^{(2,\text{squ})}_{q,\mu}(0)
            \big],
\end{equation}
when $n\in \mathbbm{N}$.~In particular, setting $n=1$ gives $	g^{(2,\text{th})}_{q,\mu}(0) =2$, further indicating that the fluctuation induced harmonics, while remaining classical, can nevertheless provide information about the photon statistics of the driving field.

\end{document}